\documentclass[journal,twoside,web]{ieeecolor}
\usepackage{tmi}
\usepackage{cite}
\usepackage{amsmath,amssymb,amsfonts}
\usepackage{algorithmic}
\usepackage{graphicx}
\usepackage{textcomp}

\usepackage{url}
\usepackage{color}
\usepackage{subfigure}
\usepackage{graphicx}
\usepackage{multirow}
\usepackage{booktabs} 
\usepackage[marginal]{footmisc} 
\usepackage{tipa} 
\usepackage{float} 
\usepackage{placeins}
\usepackage{bm} 
\usepackage{array} 
\usepackage{enumerate}
\usepackage{hyperref}
\usepackage{pifont} 
\usepackage{subfloat}
\usepackage{subfigure}
\usepackage{amsmath,bm}

\usepackage{diagbox}
\usepackage{booktabs}
\usepackage{amssymb}
\usepackage{balance} 
\usepackage{bbm} 

\usepackage{amsmath}
\usepackage[boxed,ruled]{algorithm2e}
\usepackage{algorithmic}

\def\BibTeX{{\rm B\kern-.05em{\sc i\kern-.025em b}\kern-.08em
    T\kern-.1667em\lower.7ex\hbox{E}\kern-.125emX}}
\begin{document}

\title{UN-SAM: Universal Prompt-Free Segmentation for Generalized Nuclei Images}
\author{Zhen Chen, Qing Xu, Xinyu Liu, and Yixuan Yuan, \IEEEmembership{Member, IEEE}
	\thanks{\quad This work was supported by the Hong Kong Research Grants Council (RGC) General Research Fund 14204321, 14220622 and the InnoHK program. \textit{(Equal contribution: Z. Chen and Q. Xu, Corresponding author: Yixuan Yuan)}}
	\thanks{\quad Z. Chen is with the Centre for Artificial Intelligence and Robotics (CAIR), Hong Kong Institute of Science \& Innovation, Chinese Academy of Sciences, Hong Kong SAR. (e-mail: zhen.chen@cair-cas.org.hk).}
        \thanks{\quad Q. Xu, X. Liu and Y. Yuan are with Department of Electronic Engineering, Chinese University of Hong Kong, Hong Kong SAR, China (e-mail: yxyuan@ee.cuhk.edu.hk).}
}
\maketitle
\begin{abstract}
In digital pathology, precise nuclei segmentation is pivotal yet challenged by the diversity of tissue types, staining protocols, and imaging conditions. Recently, the segment anything model (SAM) revealed overwhelming performance in natural scenarios and impressive adaptation to medical imaging. Despite these advantages, the reliance of labor-intensive manual annotation as segmentation prompts severely hinders their clinical applicability, especially for nuclei image analysis containing massive cells where dense manual prompts are impractical. To overcome the limitations of current SAM methods while retaining the advantages, we propose the Universal prompt-free SAM framework for Nuclei segmentation (UN-SAM), by providing a fully automated solution with remarkable generalization capabilities. Specifically, to eliminate the labor-intensive requirement of per-nuclei annotations for prompt, we devise a multi-scale Self-Prompt Generation (SPGen) module to revolutionize clinical workflow by automatically generating high-quality mask hints to guide the segmentation tasks. Moreover, to unleash the generalization capability of SAM across a variety of nuclei images, we devise a Domain-adaptive Tuning Encoder (DT-Encoder) to seamlessly harmonize visual features with domain-common and domain-specific knowledge, and further devise a Domain Query-enhanced Decoder (DQ-Decoder) by leveraging learnable domain queries for segmentation decoding in different nuclei domains. Extensive experiments prove that UN-SAM with exceptional performance surpasses state-of-the-arts in nuclei instance and semantic segmentation, especially the generalization capability in zero-shot scenarios. The source code is available at \href{https://github.com/CUHK-AIM-Group/UN-SAM}{https://github.com/CUHK-AIM-Group/UN-SAM}.
\end{abstract}


\section{Introduction}
\label{sec:introduction}

\begin{figure}[h]
  \centering
  \includegraphics[width=0.92\linewidth]{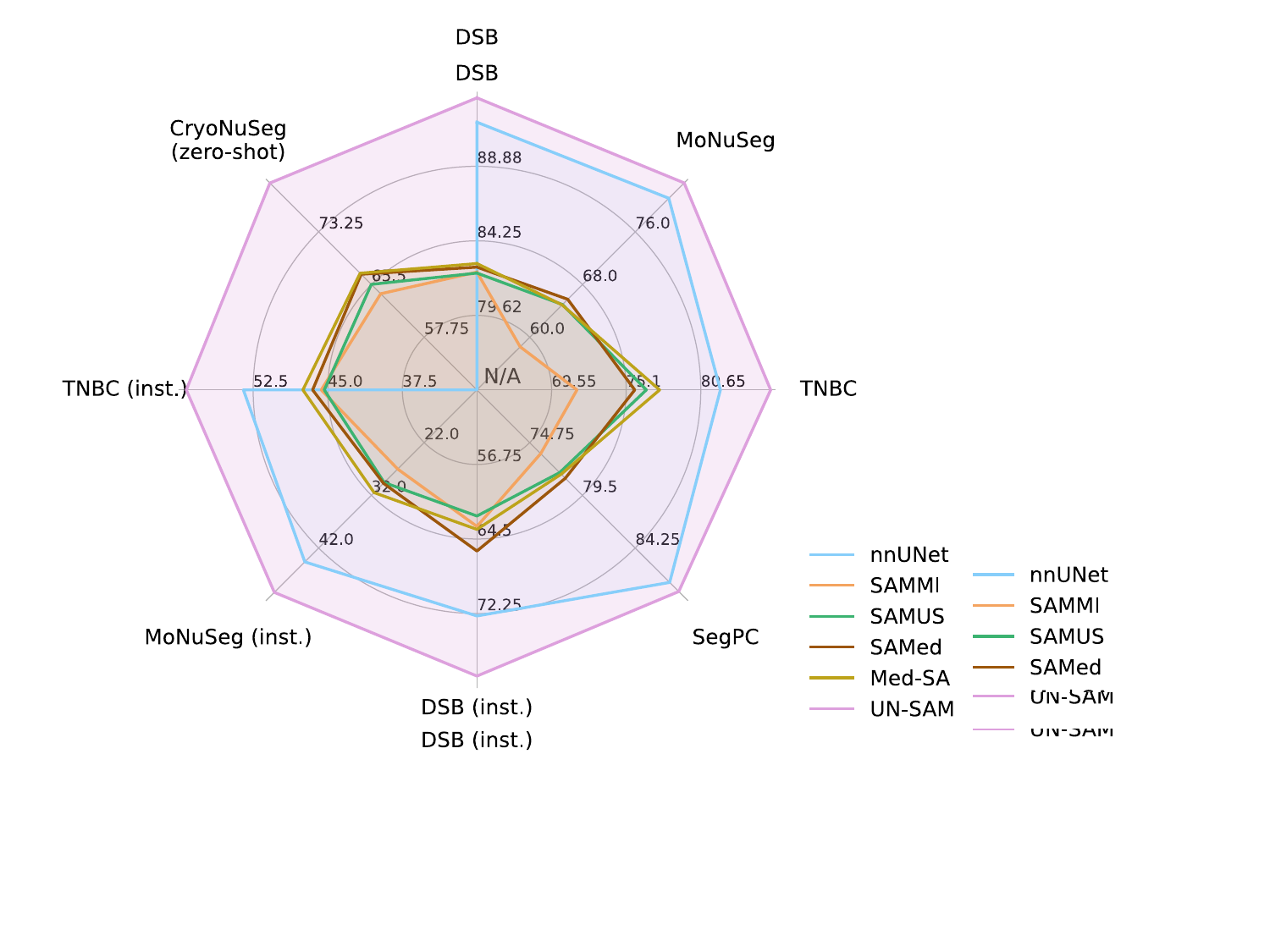}
  \caption{Performance comparison on nuclei image segmentation. The semantic segmentation is measured by Dice, and instance segmentation (marked as \textit{inst.}) is measured by Aggregated Jaccard Index (AJI).}
  \label{fig:radar}
\end{figure}

\IEEEPARstart{I}{n} the field of digital pathology, the task of nuclei image segmentation on histopathological images stands as a cornerstone for morphological quantification and tumor grade assessment \cite{greenwald2022whole}. Despite its significance, this task poses substantial difficulties to nuclei image segmentation tasks \cite{caicedo2019nucleus,gupta2023segpc}, arising from the highly dense packing of cell nuclei within tissue samples. In particular, instance segmentation methods designed for natural images \cite{he2017mask, cheng2022sparse} rely on region proposals and struggle to discriminate adjacent nuclei, thus compromising the precise segmentation of individual nucleus.

In this challenging context, classical nuclei image segmentation algorithms \cite{graham2019hover,chen2023cpp, he2021cdnet,nam2023pronet,he2023toposeg} are investigated through tailored network structures and supervision strategies targeting the characteristics of cell nuclei. These studies perform with high accuracy under fully supervised conditions in homogeneous datasets, and play an integral role in advancing the field. However, these methods require carefully crafted post-processing to identify nuclei instances, which are vulnerable to intensive hyper-parameter search. More seriously, their limitations become apparent when faced with the inherent heterogeneity of tissue types, staining protocols, and imaging conditions encountered in routine clinical practice. This variability poses significant challenges to developing segmentation models with the necessary generality and accuracy across different datasets.

The recent advent of the segment anything model (SAM) \cite{kirillov2023_sam} has ushered in a new era of segmentation, and holds promise in addressing these issues with versatile capabilities that extend beyond traditional constraints. The remarkable efficacy of SAM in natural image scenarios has been well validated to showcase its robustness and adaptability across a variety of scenarios. On this basis, SAM has begun to penetrate the medical imaging community and revealed its capability in the segmentation of medical images with the potential for diverse scenarios \cite{cheng2023sam_med2d,zhang2024segment_sam_survery,ma2024sam_majun_nc,huang2024segment}. Additionally, preliminary investigations have indicated that SAM can adapt to the unique challenges posed by medical data, though further refinement is necessary for clinical deployment \cite{lin2023samus,zhang2023customized_sam_liudong,wu2023medsa_jinyueming}. These studies collectively underscore the potential of SAM to revolutionize medical image segmentation tasks across diverse scenarios.

Despite the progress made by existing medical SAM studies \cite{lin2023samus,zhang2023customized_sam_liudong,wu2023medsa_jinyueming}, the adaptation of SAM to nuclei segmentation is hampered by two significant obstacles, including the reliance on manual annotations for segmentation prompt and the challenge of generalizing across diverse nuclei images. Firstly, one limitation of medical SAM algorithms is the reliance on manual annotations, \textit{e.g.}, the point-based prompt and bounding box-based prompt, to guide the segmentation decoding of the target objects \cite{ma2024sam_majun_nc,huang2024segment}. Although SAM provides an automatic mode to generate the bounding box-based prompt by filtering a sliding window of the input image \cite{kirillov2023_sam}, current studies \cite{cheng2023sam_med2d,huang2024segment} have demonstrated that such simple automatic prompts perform poorly in medical image segmentation, and even the simple manual annotations (\textit{e.g.}, using only a pair of positive and negative points) are difficult to ensure that medical SAM achieves satisfactory segmentation predictions. In contrast, most existing medical SAM studies \cite{ma2024sam_majun_nc,lin2023samus} rely on more manual annotations, by labeling extra points or bounding boxes as the prompt. The extensive use of manual annotations interrupts clinical workflows, and makes the process resource-intensive and typically impractical in clinical settings \cite{xu2023sppnet},  especially for nuclei images with massive cells that are extremely costly. Therefore, the ideal SAM algorithm in nuclei segmentation tasks should eliminate the need for manual annotation, and automatically achieve accurate nuclei segmentation based on the knowledge of foundation models.

Furthermore, the heterogeneity of nuclei segmentation tasks, which encompass various domains such as different tissue types, staining protocols, and imaging equipment \cite{caicedo2019nucleus}, as well as distinct task requirements like semantic and instance segmentation, necessitates the segmentation algorithm with robust generalization capabilities. However, existing medical SAM algorithms \cite{ma2024sam_majun_nc,huang2024segment} primarily rely on their inherent capacities to transfer general knowledge to medical imaging, \textit{e.g.}, directly utilizing the image encoder of SAM \cite{ma2024sam_majun_nc} and fine-tuning its encoder on downstream medical data \cite{lin2023samus,zhang2023customized_sam_liudong,wu2023medsa_jinyueming}. As such, these strategies, by simply utilizing the knowledge SAM, may be relatively difficult to deliver high performance across different medical datasets, especially for histopathology nuclei analysis where the diversity and complexity of images are particularly pronounced \cite{caicedo2019nucleus}. As illustrated in Fig.~\ref{fig:radar}, these medical SAMs \cite{huang2024segment,lin2023samus,zhang2023customized_sam_liudong,wu2023medsa_jinyueming} cannot perform satisfactorily on the nuclei segmentation with diverse scenes, and is even inferior to classical medical segmentation networks \cite{isensee2021nnu} with the same training data. Therefore, to enhance the generalization capability in such diverse settings, the SAM model should be improved with tailored designs for nuclei segmentation tasks.

To address these two aforementioned challenges in nuclei segmentation, we propose the Universal Prompt-Free SAM (UN-SAM) to achieve accurate and automatic semantic segmentation and instance segmentation with remarkable generalization capabilities. Specifically, to eliminate the requirement of the labor-intensive nuclei annotations for prompt, we devise a multi-scale Self-Prompt Generation (SPGen) module to automatically generate high-quality mask hints, guiding segmentation decoding of the UN-SAM. As such, the SPGen module can streamline the clinical workflow and make high-throughput analysis feasible without the labor-intensive process of manually crafted prompts. Moreover, to unleash the generalization capability of SAM across a variety of nuclei images, we devise a Domain-adaptive Tuning Encoder (DT-Encoder) for UN-SAM to seamlessly harmonize visual features with domain-common and domain-specific knowledge, and further devise a tailored Domain Query-enhanced Decoder (DQ-Decoder) for UN-SAM, by leveraging learnable domain queries to distinguish nuclei types and regions in different domains of complex semantic and instance segmentation. In Fig.~\ref{fig:radar}, the comparison on diverse nuclei image datasets proves that the proposed UN-SAM achieves superior generalization on different datasets, with a remarkable performance advantage over classical medical segmentation method \cite{isensee2021nnu} and medical SAMs \cite{huang2024segment,lin2023samus,zhang2023customized_sam_liudong,wu2023medsa_jinyueming}.

The contributions of this work are summarized as follows:
\begin{itemize}   
\item We propose a prompt-free UN-SAM framework to provide an automatic solution for nuclei instance and semantic segmentation with remarkable generalization capabilities across diverse datasets.

\item We devise a multi-scale SPGen module to autonomously generate high-quality mask hints and effectively guide the UN-SAM decoder, thereby eliminating the requirement of the labor-intensive nuclei annotations for prompt.

\item We devise a DT-Encoder to harmonize visual features with domain-common and domain-specific knowledge, and a DQ-Decoder to leverage domain queries for segmentation decoding. These two designs enable our UN-SAM to generalize across different nuclei domains. 

\item We conduct extensive experiments on diverse nuclei image datasets, and our UN-SAM outperforms state-of-the-art nuclei segmentation methods and medical SAMs, with remarkable zero-shot generalization performance.
\end{itemize}

\section{Related Work}
\subsection{Nuclei Image Segmentation}
The nuclei image segmentation is crucial for histopathology image analysis and can benefit pathologists in rendering precise diagnoses \cite{gupta2023segpc}. Existing studies can be categorized into nuclei semantic segmentation and nuclei instance segmentation. For the semantic segmentation that aims to identify the type and spatial region of nuclei, the early studies utilized the U-Net structure \cite{ronneberger2015u} to achieve automatic nuclei segmentation. To improve the network capability in segmentation, \cite{alom2019recurrent, chen2021transunet, schlemper2019attention, zhou2019unet++,lin2022contrans} further enhanced U-Net derivatives through recurrent and attention mechanisms. Particularly, ConTrans \cite{lin2022contrans} adopted a dual attention encoder to capture both global and local information of nuclei details, which efficiently recognized nuclei with different shapes in histopathology images.

In the realm of instance segmentation that further identifies each nucleus \cite{kumar2017dataset,naylor2018segmentation}, existing studies \cite{graham2019hover,chen2023cpp, he2021cdnet,nam2023pronet,he2023toposeg} has made notable strides through tailored network structures and supervision strategies targeting the characteristics of cell nuclei. For example, HoVer-Net \cite{graham2019hover} utilized horizontal and vertical distance maps to discern the boundaries of individual nuclei within histopathology images. To refine segmentation outputs, CDNet \cite{he2021cdnet} utilized directional feature maps and PROnet \cite{nam2023pronet} leveraged offset maps, to enhance the delineation of nuclei boundaries. The CPP-Net \cite{chen2023cpp} forged a different path by generating complementary boundary and distance masks for each nucleus, facilitating the separation of nuclei via the integration of these masks. Despite the progress, these nuclei instance segmentation methods \cite{graham2019hover,chen2023cpp, he2021cdnet,nam2023pronet,he2023toposeg} demand complex post-processing, and necessitate training on each nuclei image dataset, which are difficult to generalize to unseen nuclei image domains. In contrast, as a universal and prompt-free segmentation framework, our UN-SAM transcends dataset-specific limitations, and reveals superior generalization across a variety of nuclei image domains.

\begin{figure*}[t]
  \centering
  \includegraphics[width=0.95\linewidth]{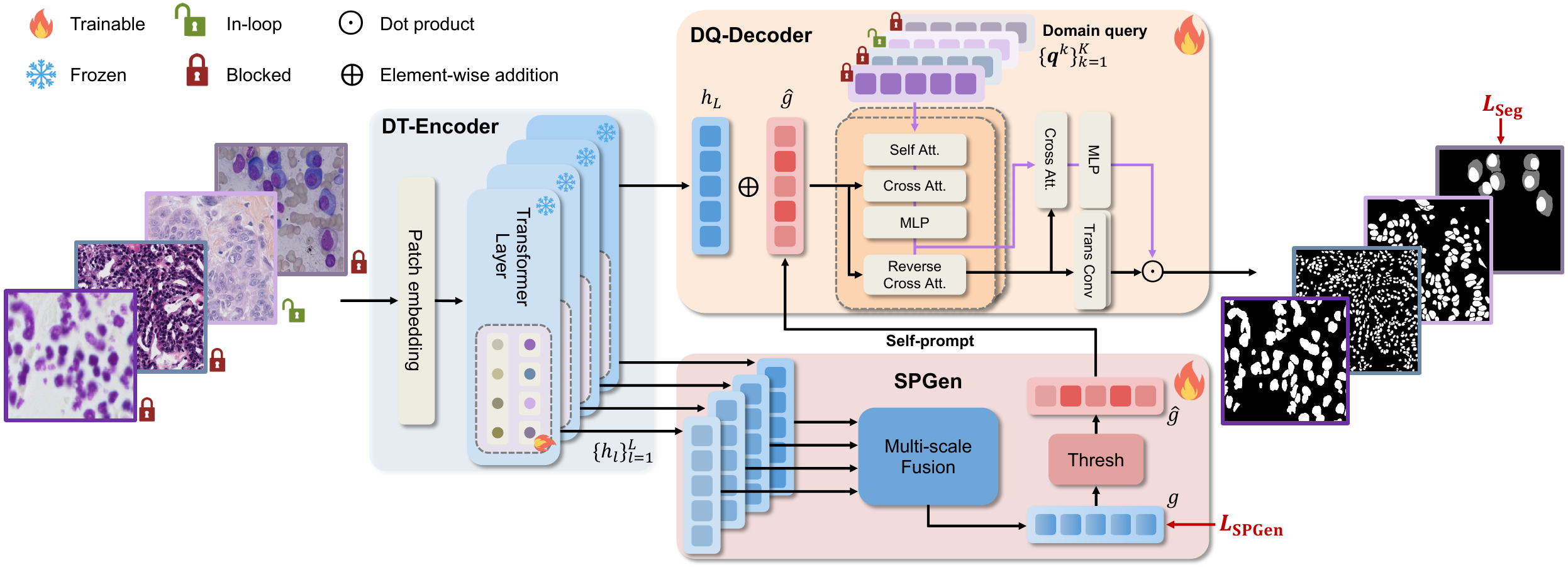}
  \caption{The overview of the proposed UN-SAM for nuclei image segmentation, consisting of DT-Encoder, SPGen and DQ-Decoder. For ease of understanding, we elaborate on the case of UN-SAM with four nuclei image domains. Our UN-SAM can achieve superior generalization performance on these domains without the need for manual annotations.}
  \label{fig:un-sam}
\end{figure*}

\subsection{The SAM in Medical Imaging}
The segment anything model (SAM) \cite{kirillov2023_sam} has well revealed the advantage in image segmentation across a variety of scenarios. By leveraging both sparse (\textit{e.g.}, point, box, and text) and dense (\textit{e.g.}, mask) prompts, SAM benefits from a robust feature extraction capability \cite{xiong2023efficientsam}, enabling it to perform zero-shot generalization for diverse image segmentation tasks. To migrate the powerful segmentation capabilities of SAM to downstream scenarios, existing works adopted different fine-tuning strategies, including directly fine-tuning the image encoder \cite{shui2023unleashing_sam_yanglin} or mask decoder \cite{ma2024sam_majun_nc}. Considering the huge amount of SAM parameters, the parameter efficient fine-tuning (PEFT) has become a hot research topic, such as the low-rank adaptation (LoRA) \cite{hu2022lora} and adapter \cite{houlsby2019parameter} techniques. The recent Conv-LoRA \cite{anonymous2024convolution} introduced a PEFT strategy for tuning SAM, and facilitated feature representation with inductive biases by dynamically selecting the proper feature scale.

On this basis, many medical SAM works \cite{ma2024sam_majun_nc,huang2024segment,lin2023samus,zhang2023customized_sam_liudong,wu2023medsa_jinyueming} have been investigated to customize segmentation capability to medical imaging. Huang \textit{et al.} \cite{huang2024segment} explored the capability of vanilla SAM \cite{kirillov2023_sam} with different types of prompts on medical image segmentation, and further fine-tuned the SAM with point and bounding box prompts on large-scale medical image datasets, which surpasses the state-of-the-art performance on 45 public datasets. The MedSAM \cite{ma2024sam_majun_nc} collected a diverse medical image segmentation dataset dominated by CT and MRI scans for tuning SAM with bounding box prompts. Furthermore, the SAMUS \cite{lin2023samus} adopted the PEFT strategy and integrated additional convolutional network embeddings with adapter modules to refine feature representations of the image encoder for downstream datasets. For the surgical images, the SurgicalSAM \cite{jieboluo2024surgicalsam} introduced the class prototypes and designated target class to guide the segmentation with the category information. In general, most medical SAMs either demand huge computational resources in fine-tuning \cite{shui2023unleashing_sam_yanglin} or rely on manual annotations for prompt during inference \cite{ma2024sam_majun_nc,huang2024segment,lin2023samus,zhang2023customized_sam_liudong,wu2023medsa_jinyueming}, which is impractical for nuclei images containing hundreds or thousands of nuclei in each image. Our UN-SAM endeavors to resolve this bottleneck with tailored efficient training and automatic prompt generation mechanisms, thereby streamlining the segmentation process for nuclei images.

In the field of nuclei segmentation, All-in-SAM \cite{cui2023all_in_sam} utilized high-frequency image information extracted by Fourier transform to tune the prompt encoder and mask decoder of SAM. With the centroid of each nucleus as prompt, the SPPNet \cite{xu2023sppnet} calculated the surrounding neighborhood points of centroids as auxiliary prompts to facilitate the segmentation. Different from existing methods that targeted a single nuclei domain \cite{cui2023all_in_sam,xu2023sppnet,shui2023unleashing_sam_yanglin}, our UN-SAM fully leverages the knowledge of foundation models, and achieves generalization on varying nuclei domains without any manual prompts.

\section{Methodology}
\subsection{Overview of UN-SAM}
As illustrated in Fig. \ref{fig:un-sam}, we present the UN-SAM to provide automatic segmentation with remarkable generalization across a variety of nuclei images. Given nuclei images from the $k$-th domain, we first utilize the Domain-adaptive Tuning Encoder (DT-Encoder) to generate domain knowledge-enhanced image embeddings. Then, these embeddings are delivered to the multi-scale Self-Prompt Generation (SPGen) module, which autonomously produces a set of self-generated prompt tokens to guide nuclei segmentation. Following this, the Domain Query-enhanced Decoder (DQ-Decoder) leverages the query embedding of the corresponding domain to accurately predict the segmentation mask. 

In general, this novel UN-SAM, comprising the DT-Encoder and DQ-Decoder, is specifically designed to tackle the challenges of model generalization across varying domains in nuclei image segmentation. Meanwhile, the SPGen module eliminates the demand of labor-intensive manual annotations for prompts that are typically associated with the SAM frameworks \cite{kirillov2023_sam,ma2024sam_majun_nc}. Therefore, these key modules form our UN-SAM that enhances the generalization and applicability to a wide array of nuclei image segmentation.

\subsection{Domain-adaptive Tuning Encoder}
Recent studies have attributed the remarkable segmentation capabilities of SAM to its large-capacity image encoder \cite{zhang2024segment_sam_survery, xiong2023efficientsam}. When adapting SAM to medical images, existing studies mainly utilized the pre-trained SAM encoder directly \cite{ma2024sam_majun_nc} or fine-tuned it on downstream datasets \cite{lin2023samus,zhang2023customized_sam_liudong,wu2023medsa_jinyueming}. However, these studies that rely entirely on a set of image encoder parameters are suboptimal for the complex task of nuclei segmentation, especially considering the heterogeneity of the nuclei image domains. To extract discriminative visual features for nuclei segmentation across different domains, we propose a Domain-adaptive Tuning Encoder (DT-Encoder) that leverages a learnable set of domain-common and domain-specific embeddings for downstream fine-tuning with a tailored strategy, facilitating the UN-SAM with superior segmentation performance.

\begin{figure}[t]
  \centering
  \includegraphics[width=0.85\linewidth]{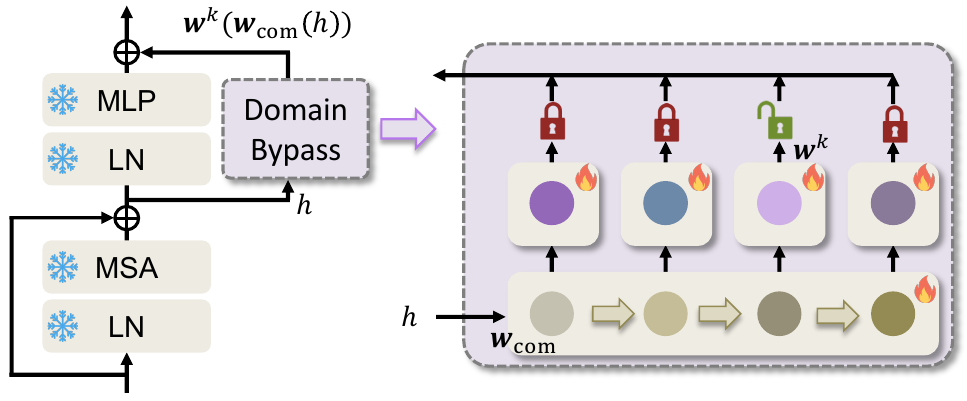}
  \caption{The illustration of the transformer layer with the domain bypass in the  DT-Encoder. In the domain bypass, the domain-common and domain-specific embeddings process image embeddings in sequence.}
  \label{fig:spgen}
\end{figure}

Specifically, the DT-Encoder follows the ViT \cite{dosovitskiy2020image} architecture of the SAM encoder \cite{kirillov2023_sam}, consisting of $L$ transformer layers. To enhance the transformer layer with the domain-adaptive capability, we devise the domain bypass to process the image embedding $h$ after the multi-head self-attention (MSA) operation, as illustrated in Fig.~\ref{fig:spgen}. For each transformer layer, we introduce a learnable set of domain-common embedding $\bm{w}_{\rm com}$ and domain-specific embeddings $\{\bm{w}^{k}\}_{k=1}^{K}$ for the domain bypass, where $K$ is the number of nuclei domains. Note that we only update the learnable $\bm{w}_{\rm com}$ and $\{\bm{w}^{k}\}_{k=1}^{K}$ in the UN-SAM fine-tuning, and keep the original parameters in SAM frozen.

In the domain bypass, we first leverage the domain-common embeddings $\bm{w}_{\rm com}$ to process the image embedding $h$ using the common knowledge of $K$ nuclei domains, and then enhance the image embedding $h$ with the corresponding domain-specific knowledge, \textit{e.g.}, using the domain-specific embedding $\bm{w}^{k}$ for $k$-th nuclei domain. After that, the output of the domain bypass is added to enhance the image embedding in the residual, as follows:
\begin{equation}
    h \gets  \bm{w}^k( \bm{w}_{\rm com} (h)) + {\rm MLP}(h),
\end{equation}
where ${\rm MLP}$ is the multi-layer perceptron (MLP) after the layer normalization (LN) operation. In this way, the domain bypass in each transformer layer enables the generalization of the DT-Encoder across a variety of nuclei images by introducing a slight amount of learnable parameters.

Furthermore, to provide the DT-Encoder with adequate nuclei image knowledge while freezing SAM parameters, we devise a domain inheritance strategy for domain-common embedding to fully exploit discriminative features for nuclei image segmentation. When our UN-SAM is trained on different domains sequentially, we let the domain-common embedding of the next domain $\bm{w}_{\rm com}^{k+1}$ inherit from the previous domain $\bm{w}_{\rm com}^{k}$ to maintain continuous learning on nuclei segmentation, as follows:
\begin{equation}\label{eq:domain_inheritance}
    \bm{w}_{\rm com}^{k+1} \gets \bm{w}_{\rm com}^{k}.
\end{equation}
Therefore, the DT-Encoder can exploit nuclei image knowledge adequately and make the fine-tuning efficient. In contrast, the domain-specific embedding is used for the corresponding domain of nuclei images, which makes the DT-Encoder adaptive to different nuclei domains. In this way, the proposed DT-Encoder can unleash the generalization capability of UN-SAM with both domain-common and domain-specific knowledge.

\subsection{Multi-scale Self-prompt Generation}
The standard SAM \cite{kirillov2023_sam} and medical SAM \cite{huang2024segment, ma2024sam_majun_nc} demand manual annotations as prompts during image segmentation, \textit{e.g.}, the point prompt and bounding box prompt. In practice, manually labeling points and even bounding boxes of medical images are time-consuming and expensive in clinical scenarios, especially for histopathology images with numerous nuclei. To address this issue, we propose a multi-scale Self-Prompt Generation (SPGen) module that can automatically provide a set of high-quality self-prompt with multi-scale knowledge to facilitate nuclei segmentation tasks.

As depicted in Fig.~\ref{fig:un-sam}, the proposed SPGen module first takes the image embeddings $\{h_l\}_{l=1}^{L}$ from different layers of DT-Encoder as the input, and performs the multi-scale fusion to generate the multi-scale image embeddings ${h}_{\rm ms}$. To achieve this, we adopt convolutional layer heads with different strides on $\{h_l\}_{l=1}^{L}$ to produce the image embeddings at different scales, and then apply the Feature Pyramid Network (FPN) \cite{ali2021xcit} to integrate multi-level features from top to down. As such, the generated multi-scale image embeddings $h_{\rm ms}$ are capable to perceive nuclei with different sizes and exploit discriminative representations for the segmentation hint.

After that, to facilitate the decoding of nuclei segmentation, the SPGen module generates the self-prompt by filtering out high-quality foreground regions. First, we apply a $1\times 1$ convolutional head on the multi-scale image embedding $h_{\rm ms}$ to predict the foreground regions. The foreground segmentation is optimized by the binary cross-entropy loss, as follows:
\begin{equation} \label{eq:spgen}
L_{\rm SPGen}= -\frac{1}{N} \sum_{n=1}^{N}y_{n} \log (\sigma(g_{n}))
 +(1-y_{n}) \log (1-\sigma(g_{n})),
\end{equation}
where $y_n$ is the $n$-th token of ground truth, and $N$ is the token numbers of image embeddings. The sigmoid function $\sigma$ processes $g_{n}$ into the predicted foreground probability, and a higher score means the corresponding token plays a more important role in nuclei prediction. To avoid the generation of noisy hints and reduce the false positive error in mask prediction, we dynamically retain the tokens with high probability and filter out low-confidence tokens prone to misclassification, as follows:
\begin{equation}
\hat{g}_n= \begin{cases}g_n, & \text { if } \sigma\left(g_n\right) \geqslant \tau \\ 0, & \text { otherwise }\end{cases}
\end{equation}
where $\tau$ is a threshold to determine the high-quality foreground tokens. In this way, the proposed SPGen module can produce a set of high-confidence self-prompt tokens $\hat{g}=\{\hat{g}_n\}_{n=1}^{N}$ to promote the nuclei segmentation, and eliminate the need for UN-SAM on manual annotations.

\subsection{Domain Query-enhanced Decoder}
The decoder in the standard SAM \cite{kirillov2023_sam} and medical SAM \cite{huang2024segment,ma2024sam_majun_nc} adopted the learnable output token as the query to generate the corresponding segmentation mask. However, it is challenging to leverage the same set of query embeddings to process nuclei images in different domains. To further promote the generalization of UN-SAM across a variety of domains, we propose a Domain Query-enhanced Decoder (DQ-Decoder) to generate accurate segmentation masks with domain-specific query embeddings.

To enhance the image embeddings with foreground hints, the DQ-Decoder takes the self-prompt $\hat{g}$ from the SPGen module and the image embedding $h_L$ from the DT-Encoder as input, and performs element-wise addition, as $f=h_{L}+\hat{g}$. Meanwhile, to facilitate UM-SAM to segment a variety of nuclei images, we set a group of domain query embeddings $\{\bm{q}^k\}_{k=1}^{K}$, where $\bm{q}^k\in \mathbb{R}^{N\times d}$ is for $k$-th domain, and $\bm{q}^k$ has the same shape as $f$. As such, the DQ-Decoder first performs the self-attention layer based on domain query embeddings $\bm{q}^k$, followed by the cross-attention layer with the hint-enhanced image embedding $f$ to update the domain query embeddings $\bm{q}^k$, as follows: 
\begin{equation}
    \bm{q}^k \gets \mathrm{softmax}(\frac{\bm{q}^{k} \cdot {(f + \psi)}^{T}}{\sqrt{d}})\cdot f+\bm{q}^k,
\end{equation}
where $\cdot$ is the matrix multiplication, and the positional encoding $\psi$ is inserted to improve the dependence between geometric location and type. After the MLP, the DQ-Decoder further adopts the reverse cross-attention layer to update the image embedding $f$ with the updated domain query embeddings $\bm{q}^k$, as follows:
\begin{equation}
    f \gets \mathrm{softmax}(\frac{(f+ \psi) \cdot (\bm{q}^k)^{T}}{\sqrt{d}})\cdot \bm{q}^k+f,
\end{equation}
Following the standard SAM \cite{kirillov2023_sam}, we repeat the decoder block of the above processes twice to obtain updated domain queries and image embedding.

\IncMargin{-1em} 
\begin{algorithm}[tbp!]
\caption{{UN-SAM pipeline across varying domains.}} 
\SetKwInOut{Input}{\textbf{Input}}\SetKwInOut{Output}{\textbf{Output}} 
\hspace*{0.02in}\Input{The network structure: $\bm{E}$, $\bm{S}$, $\bm{D}$;\\
    \hspace*{0.02in}Learnable $\bm{w}_{\rm com}$, $\{\bm{w}^{k}\}_{k=1}^{K}$, $\{\bm{q}^{k}\}_{k=1}^{K}$;\\
    \hspace*{0.02in}Domains: $K$, Epochs: $M$;}
    \hspace*{0.02in}\Output{The well-trained UN-SAM.}

\vspace{2.0pt}
\begin{algorithmic}[1]
\STATE Initialize the model parameters $\bm{E}$, $\bm{S}$, $\bm{D}$ for all domains;\\
\FOR{{\rm domain} $k=1$ to $K$}
\FOR{{\rm epoch} $m=1$ to $M$}
\STATE Perform DT-Encoder: $\{h_{l}\}_{l=1}^{L}$ $\gets$ $\bm{E}(x;\bm{w}_{\rm com},\bm{w}^{k})$;\\
\STATE Perform SPGen: $\hat{g}$ $\gets$ $\bm{S}(\{h_{l}\}_{l=1}^{L})$;\\
\STATE Perform DQ-Decoder: $\hat{y}$ $\gets$ $\bm{D}(h_{L}, \hat{g}; \bm{q}^{k})$;\\
\STATE Compute $L_{\rm SPGen}$ in Eq.~\eqref{eq:spgen};\\
\STATE Compute $L_{\rm Seg}$ in Eq.~\eqref{eq:algorithm};\\
\STATE Optimize $\bm{S}$ and $\bm{D}$ with learnable $\bm{w}_{\rm com}$, $\bm{w}^{k}$, $\bm{q}^{k}$.\\
\ENDFOR
\STATE Update $\bm{w}_{\rm com}$ in $\bm{E}$: $\bm{w}_{\rm com}^{k+1}$ $\gets$ $\bm{w}_{\rm com}^{k}$ in Eq.~\eqref{eq:domain_inheritance};\\
\ENDFOR
\end{algorithmic}
\end{algorithm}\label{algorithm}
\DecMargin{-1em}

After that, we upscale the domain query-enhanced image embedding $f$ with two transposed convolutional layers into the targeted resolution. Meanwhile, we update the domain query with a cross-attention layer with the image embedding, followed by a MLP to adjust the channel dimension the same as the upscaled image embedding. Finally, the DQ-Decoder generates the segmentation prediction by performing the dot product between the upscaled image embedding and the $k$-th domain query. In this way, the domain information in the DQ-Decoder does not interfere with each other, which further improves the generalization of the proposed UN-SAM.

\subsection{Optimization Pipeline}
We summarize the training pipeline of UN-SAM across varying nuclei domains in Algorithm \ref{algorithm}. We first initialize the UN-SAM including the DT-Encoder ($\bm{E}$), SPGen module ($\bm{S}$) and DQ-Decoder ($\bm{D}$), and particularly we utilize the pre-trained SAM \cite{kirillov2023_sam} to initialize corresponding parameters in the DT-Encoder and keep them frozen. The UN-SAM is optimized on varying nuclei domains in sequence, and at the end of each domain, the $\bm{w}_{\rm com}$ inherits the parameters from the previous domain ones. For each batch of $k$-th domain nuclei images, the DT-Encoder generates the domain-adaptive features $\{h_l\}_{l=1}^{L}$, then the SPGen module provides the self-prompt $\hat{g}$ as the hint for segmentation, and finally the DQ-Decoder generates the segmentation prediction $\hat{y}$. The predicted segmentation mask $\hat{y}$ on each domain is supervised by the weighted combination of focal loss $L_{\rm focal}$ and dice loss $L_{\rm dice}$ \cite{kirillov2023_sam}, as follows:
\begin{equation} \label{eq:algorithm}
    L_{\rm Seg} = \lambda L_{\rm focal} + (1-\lambda) L_{\rm dice},
\end{equation}
where $\lambda$ is the coefficient to balance these two loss terms. By optimizing $L_{\rm Seg}$ and $L_{\rm SPGen}$, our UN-SAM achieves accurate segmentation with superior generalization across different domains, without the need for manual annotations.

\begin{table*}[!t]
    \centering
    \small
    \setlength\tabcolsep{5pt}
    \caption{Comparison with State-of-the-arts on Nuclei Instance Segmentation.}
    {\scalebox{0.82}{
    \begin{tabular}{l|c|cccccccccccccc}
    \toprule
    \multirow{2}{*}{Methods} & Manual & \multicolumn{4}{c}{DSB} & & \multicolumn{4}{c}{ MoNuSeg } & & \multicolumn{4}{c}{ TNBC }\\
    \cline{3-6} \cline{8-11} \cline{13-16 }
     & Prompt & AJI & DQ & SQ & PQ & & AJI & DQ & SQ & PQ & & AJI & DQ & SQ & PQ\\
    \midrule
    U-Net \cite{ronneberger2015u} & \multirow{7}{*}{\ding{56}} & 71.53 & 81.37 & 83.42 & 68.89 & & 34.98 & 50.94 & 68.59 & 34.96  &  & 50.52 & 68.02 & 74.24 & 50.52 \\
    nnU-Net \cite{isensee2021nnu} &  & 74.57 & 84.49 & 84.07 & 72.11 & & 45.22 & 65.12 & 71.78 & 48.81 &  & 56.48 & 75.77 & 77.97 & 59.30\\
    Mask-RCNN \cite{he2017mask}&  & 71.80 & 82.07 & 83.11 & 69.23 & & 36.15 & 52.17 & 70.03 & 37.46 &  & 51.04 & 67.33 & 75.01 & 50.83 \\
    SparseInst \cite{cheng2022sparse} & & 72.50 & 83.67 & 83.93 & 71.35 & & 40.19 & 55.89 & 70.01 & 40.35 & & 53.48 & 68.26 & 76.04 & 53.94\\
    HoVer-Net \cite{graham2019hover} & &  74.91 & 85.23 & 84.13 & 73.45 & & 46.76 & 66.37 & 70.92 & 47.58 & & 56.45 & 76.19 & 77.65 & 59.11\\
    CDNet \cite{he2021cdnet} &  &  74.87 & 84.89 & 84.30 & 73.12 & & 43.49 & 59.77 & 70.16 & 43.98 & & 54.25 & 74.28 & 76.13 & 56.85\\
    CPP-Net \cite{chen2023cpp} & & 75.54 & 86.95 & 84.58 & 74.35 & & 44.63 & 60.57 & 70.53 & 42.44 &  & 52.20 & 69.17 & 74.75 & 51.67 \\
    
    \midrule
    Vanilla SAM \cite{kirillov2023_sam} & \multirow{5}{*}{\ding{56}} & 22.30 & 16.17 & 49.32 & 12.96 & & 7.36 & 8.55 & 50.14 & 6.08 & & 8.42 & 8.95 & 38.92 & 5.90 \\
    SAMMI \cite{huang2024segment} &  & 63.19 & 73.93 & 72.54 & 59.90 & & 27.02 & 45.45 & 57.39 & 28.88 & & 45.65 & 57.58 & 68.88 & 41.46 \\
    SAMUS \cite{lin2023samus} &  & 62.11 & 74.14 & 72.52 & 59.97 & & 29.58 & 47.14 & 58.17 & 31.88 & & 45.35 & 58.85 & 68.06 & 41.39 \\
    SAMed  \cite{zhang2023customized_sam_liudong} &  & 65.75 & 75.19 & 74.48 & 62.58 & & 29.72 & 46.93 & 60.52 & 30.21 &  & 46.52 & 59.69 & 68.36 & 40.21 \\
    Med-SA  \cite{wu2023medsa_jinyueming} &  & 63.50 & 75.78 & 73.09 & 61.73 & & 31.49 & 47.71 & 61.31 & 32.32 &  & 47.50 & 59.67 & 70.39 & 43.32 \\
    \midrule
    Vanilla SAM \cite{kirillov2023_sam}& \multirow{5}{*}{point}  & 36.67 & 45.10 & 62.47 & 36.16 & & 14.11 & 14.64 & 53.46 & 9.32 &  & 20.54 & 24.24 & 68.07 & 16.73 \\
    SAMMI \cite{huang2024segment}&  & 75.29 & 85.82 & 84.21 & 72.82 & & 44.83 & 61.53 & 71.41 & 44.01 &  & 57.33 & 70.61 & 76.97 & 54.40 \\
    SAMUS \cite{lin2023samus}&   & 74.93 & 86.63 & 83.86 & 73.09 &  & 43.80 & 61.71 & 71.59 & 44.24 &  & 56.53 & 73.95 & 77.71 & 57.47 \\
    SAMed  \cite{zhang2023customized_sam_liudong} &  & 75.58 & 86.82 & 84.10 & 73.55 & & 44.46 & 61.64 & 71.03 & 43.81 &  & 53.77 & 67.85 & 76.23 & 51.73 \\
    Med-SA  \cite{wu2023medsa_jinyueming} &  & 74.82 & 86.49 & 84.48 & 73.59 & & 46.00 & 65.90 & 71.87 & 47.39 &  & 57.00 & 76.79 & 77.72 & 59.72\\
    \midrule
    UN-SAM (Ours) & \ding{56} & \textbf{78.75} & \textbf{92.25} & \textbf{86.79} & \textbf{78.23} & & \textbf{50.59} & \textbf{70.37} & \textbf{72.88}  & \textbf{50.86} &  & \textbf{59.48} & \textbf{78.13} & \textbf{78.64} & \textbf{61.47} \\
    \bottomrule
    \end{tabular}}}
    \label{tab_instance_sota}
\end{table*}

\begin{table*}[t]
    \centering
    \small
    \setlength\tabcolsep{5pt}
    \caption{Comparison of Generalization Capability on Nuclei Instance Segmentation.}
    {\scalebox{0.82}{
    \begin{tabular}{l|c|cccccccccccccc}
    \toprule
    \multirow{2}{*}{Methods} & Manual & \multicolumn{4}{c}{DSB} & & \multicolumn{4}{c}{ MoNuSeg } & & \multicolumn{4}{c}{ TNBC }\\
    \cline{3-6} \cline{8-11} \cline{13-16 }
     & Prompt & AJI & DQ & SQ & PQ & & AJI & DQ & SQ & PQ & & AJI & DQ & SQ & PQ\\
    \midrule
    Vanilla SAM \cite{kirillov2023_sam} & \multirow{5}{*}{\ding{56}} & 22.30 & 16.17 & 49.32 & 12.96 & & 7.36 & 8.55 & 50.14 & 6.08 & & 8.42 & 8.95 & 38.92 & 5.90 \\
    SAMMI \cite{huang2024segment} &  & 62.37 & 73.04 & 71.96 & 58.56 & & 26.13 & 44.68 & 55.97 & 27.98 & & 43.92 & 56.17 & 66.42 & 40.59 \\
    SAMUS \cite{lin2023samus} &  & 61.03 & 72.98 & 71.26 & 58.74 & & 27.32 & 46.91 & 56.13 & 30.27 & & 44.16 & 57.54 & 67.52 & 40.78 \\
    SAMed  \cite{zhang2023customized_sam_liudong} &  & 63.98 & 74.12 & 73.33 & 60.93 & & 28.56 & 45.81 & 59.04 & 29.97 &  & 45.37 & 58.14 & 67.96 & 39.15 \\
    Med-SA  \cite{wu2023medsa_jinyueming} &  & 63.26 & 75.16 & 72.29 & 60.68 & & 30.33 & 46.62 & 60.04 & 31.15 &  & 46.43 & 58.59 & 68.74 & 42.08 \\
    \midrule
    Vanilla SAM \cite{kirillov2023_sam}& \multirow{5}{*}{point}  & 36.67 & 45.10 & 62.47 & 36.16 & & 14.11 & 14.64 & 53.46 & 9.32 &  & 20.54 & 24.24 & 68.07 & 16.73 \\
    SAMMI \cite{huang2024segment}&  & 72.37 & 83.36 & 83.58 & 69.75 & & 41.04 & 58.42 & 70.49 & 41.21 &  & 52.04 & 63.91 & 74.84 & 53.12 \\
    SAMUS \cite{lin2023samus}&  & 72.46 & 84.52 & 83.47 & 70.12 &  & 43.15 & 61.04 & 71.39 & 43.56 &  & 55.72 & 71.24 & 76.84 & 56.07 \\
    SAMed  \cite{zhang2023customized_sam_liudong} &  & 72.81 & 85.28 & 83.35 & 71.36 & & 42.57 & 60.85 & 70.76 & 41.99 &  & 51.95 & 68.30 & 75.23 & 50.16 \\
    Med-SA  \cite{wu2023medsa_jinyueming} &  & 73.14 & 85.53 & 83.38 & 71.80 & & 44.24 & 63.44 & 71.04  & 45.11 &  & 56.19 & 73.07 & 77.24 & 56.91\\
    \midrule
    UN-SAM (Ours) &  \ding{56} & \textbf{75.56} & \textbf{89.18} & \textbf{84.67} & \textbf{76.12} & & \textbf{49.37} & \textbf{69.16} & \textbf{71.74}  & \textbf{49.52} &  & \textbf{59.22} & \textbf{77.91} & \textbf{78.60} & \textbf{61.28} \\
    \bottomrule
    \end{tabular}}}
    \label{tab_instance_generalized}
\end{table*}

\section{Experiment}
\subsection{Datasets and Implementations}
\subsubsection{Datasets}
To validate the effectiveness of the proposed UN-SAM, we adopt four nuclei segmentation datasets, including the DSB \cite{caicedo2019nucleus}, MoNuSeg \cite{kumar2017dataset}, TNBC \cite{naylor2018segmentation}, SegPC \cite{gupta2023segpc} datasets, to perform comprehensive cross-domain comparisons, and one additional CryoNuSeg \cite{mahbod2021cryonuseg} dataset to evaluate zero-shot generalization capability. The details are as follows:

\noindent \textbf{DSB} \cite{caicedo2019nucleus} dataset, derived from the 2018 Data Science Bowl challenge, contains 670 microscopic slides of different image types across diverse cell lines, imaging conditions, and staining protocols. The image size varies from $256\times 256$ to $696\times 520$.

\noindent \textbf{MoNuSeg} \cite{kumar2017dataset} is a nuclei segmentation dataset collected from multiple organs, including the breast, liver, kidney, prostate, bladder, colon and stomach. It contains 44 H\&E stained images with the resolution of $1,000\times 1,000$.

\noindent \textbf{TNBC} \cite{naylor2018segmentation} is a nuclei segmentation dataset from triple-negative breast cancer patients, and contains 50 histopathology images of $512\times 512$ resolution captured at 40$\times$ magnification.

\noindent \textbf{SegPC} \cite{gupta2023segpc} is a multiple myeloma plasma cell segmentation dataset, and contains 498 nuclei images annotated with two categories of the cytoplasm and nucleus. Each image is either $2,040\times1,536$ or $2,560\times 1,920$.

\noindent \textbf{CryoNuSeg} \cite{mahbod2021cryonuseg} is a cryosectioned nuclei segmentation dataset of H\&E stained tissues from 10 different organs, and contains 30 slides of $512\times 512$ captured at 40$\times$ magnification. We keep this dataset invisible from model training to evaluate zero-shot generalization capability.

\subsubsection{Implementation Details}
We perform all experiments on a single NVIDIA A800 GPU using PyTorch. For a fair comparison, we implement all nuclei segmentation methods with the same training settings and configurations, where all SAM models use ViT-H \cite{dosovitskiy2020image} structure as the image encoder. We perform the optimization using Adam with the batch size of 4 for 30 epochs. The learning rate is initialized as $1\times10^{-4}$ and adjusted using the exponential decay strategy with the factor as $0.98$. For our UN-SAM, we set the foreground threshold $\tau$ as $0.95$ in the SPGen, and the loss coefficient $\lambda$ as $0.8$ during the training. To generate instance segmentation predictions, we use the connectedComponents function in OpenCV to identify each nucleus from semantic segmentation masks, without bells and whistles. In the comparison, the medical SAM methods \cite{huang2024segment,lin2023samus,zhang2023customized_sam_liudong,wu2023medsa_jinyueming} is fine-tuned and evaluated under the same dataset protocol as our UN-SAM. We implement two modes of these medical SAM methods, including the automatic mode without manual annotations \cite{kirillov2023_sam} and the prompt mode using the centroid of each nucleus instance as point prompts.

\subsubsection{Evaluation Metrics}
To perform the comprehensive evaluation of nuclei segmentation, we adopt diverse metrics in terms of semantic segmentation and instance segmentation.
For nuclei semantic segmentation, we select the Dice coefficient, mean intersection over union (mIoU), F1 score, and Hausdorff distance (HD). For nuclei instance segmentation, we compare the performance with four metrics, including the aggregated Jaccard index (AJI), detection quality (DQ), segmentation quality (SQ), and panoptic quality (PQ). Except for HD, higher scores for these metrics indicate better segmentation quality.

\subsection{Comparison on Nuclei Instance Segmentation}
To evaluate our UN-SAM in nuclei instance segmentation, we perform the comparison with state-of-the-art instance segmentation methods \cite{he2017mask,cheng2022sparse}, nuclei instance segmentation methods \cite{graham2019hover,he2021cdnet,chen2023cpp} and medical SAM \cite{huang2024segment, lin2023samus, wu2023medsa_jinyueming,zhang2023customized_sam_liudong} on the DSB, MoNuSeg and TNBC datasets. Except for the vanilla SAM \cite{kirillov2023_sam} as the baseline for foundation models, we guarantee fair comparisons and fine-tune all models with the same training set to adapt to nuclei images. First, we evaluate the performance of all models fine-tuned on each single nuclei domain, as illustrated in Table~\ref{tab_instance_sota}. We observe that the fine-tuned SAM methods \cite{huang2024segment, lin2023samus, wu2023medsa_jinyueming,zhang2023customized_sam_liudong} are inferior to nuclei instance segmentation methods \cite{graham2019hover,he2021cdnet,chen2023cpp} when manual prompt is unavailable. In contrast, given each nucleus centroid as the point prompt, these medical SAMs reveal the advantage of foundation models, \textit{e.g.}, Med-SA \cite{wu2023medsa_jinyueming} surpasses CPP-Net \cite{chen2023cpp} with a 1.37\% AJI increase on the MoNuSeg dataset. Remarkably, our UN-SAM, without relying on manual prompts, achieves the best performance among different domains of instance nuclei segmentation, including AJI of 78.75\%, 50.59\% and 59.48\% on the respective datasets. These comparisons prove the efficacy of our UN-SAM, showcasing its superior performance on diverse nuclei segmentation datasets without the need for manual annotation.

\begin{figure}[t]
  \centering
  \includegraphics[width=0.99\linewidth]{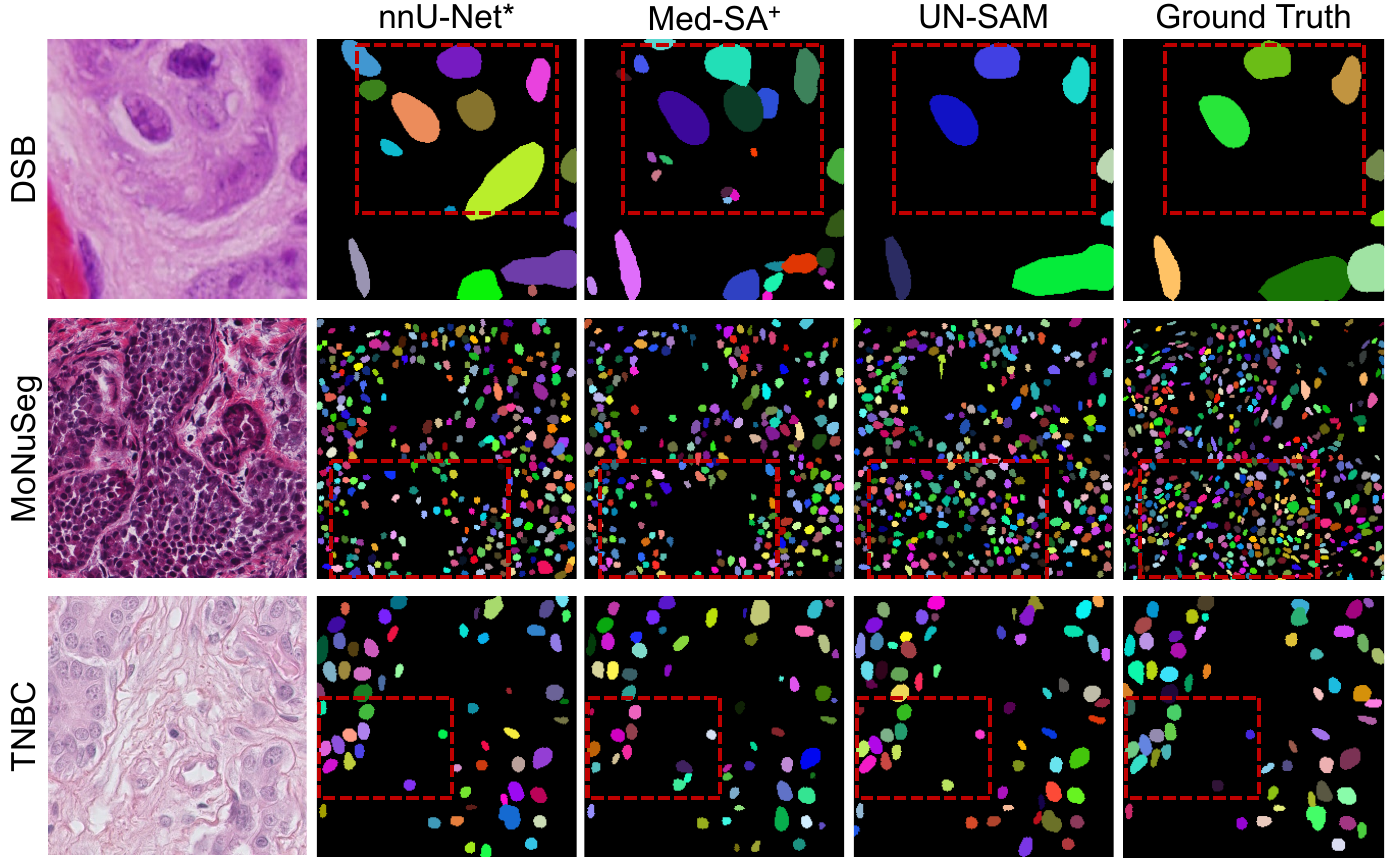}
  \caption{Visualization of generalized nuclei instance segmentation. ${+}$ indicates medical SAMs using point prompts, and $*$ indicates that classical methods are trained and tested separately for each dataset. Our UN-SAM segments more nuclei with accurate boundaries while having fewer false positives.}
  \label{fig:visual_instance}
\end{figure}

Furthermore, to evaluate the generalization capability across different nuclei domains, we compare the performance of the UN-SAM and SAM-based methods on the DSB, MoNuSeg and TNBC datasets after fine-tuning them on these three datasets sequentially, as shown in Table \ref{tab_instance_generalized}. Our UN-SAM reveals the overwhelming generalization performance of four metrics on these datasets. When manual prompts are unavailable for medical SAM, our UN-SAM achieves a significant advantage over the second-best Med-SA \cite{wu2023medsa_jinyueming}, \textit{e.g.}, a PQ increase of 15.44\%, 18.37\% and 19.20\% on the DSB, MoNuSeg and TNBC datasets, respectively. Even compared with the medical SAM using nucleus centroids as prompt (denote as \textit{point}), our UN-SAM also achieves superior performance, by outperforming these medical SAM models \cite{isensee2021nnu,zhang2023customized_sam_liudong,wu2023medsa_jinyueming} with the P-value $< 0.005$ in both AJI and PQ. 
We further compare the quantitative results of our UN-SAM with the best baseline methods \cite{isensee2021nnu,wu2023medsa_jinyueming} in Fig.~\ref{fig:visual_instance}, and our UN-SAM can segment nuclei more accurately with better boundaries. In this way, these results demonstrate the significant generalization advantage of our UN-SAM over medical foundation models on different nuclei segmentation datasets.

\begin{table*}[!t]
    \centering
    \small
    \setlength\tabcolsep{5pt}
    \caption{Comparison with State-of-the-arts on Nuclei Semantic Segmentation.}
    {\scalebox{0.77}{
    \begin{tabular}{l|c|cccccccccccccccccccc}
    \toprule
    \multirow{2}{*}{Methods} & Manual & \multicolumn{4}{c}{DSB} & & \multicolumn{4}{c}{ MoNuSeg } & & \multicolumn{4}{c}{ TNBC } & & \multicolumn{4}{c}{ SegPC }\\
    \cline{3-6} \cline{8-11} \cline{13-16} \cline{18-21}
    & Prompt & Dice & mIoU & F1 & HD & & Dice & mIoU & F1 & HD & & Dice & mIoU & F1 & HD & & Dice & mIoU & F1 & HD\\
    \midrule
    U-Net \cite{ronneberger2015u} & \multirow{9}{*}{\ding{56}} & 88.16 & 81.42 & 89.49 & 32.57 & & 74.06 & 60.25 & 75.57 & 18.94 & & 80.64 & 67.62 & 81.00 & 38.39  & & 84.89 & 76.20 & 86.75 & 58.17\\ 
    Unet++ \cite{zhou2019unet++} &  & 90.48 & 83.53 & 90.96 & 28.03 & & 76.78 & 62.97 & 78.31 & 18.53 & & 81.19 & 68.44 & 81.60 & 37.70 & & 85.61 & 77.10 & 87.04 & 56.61\\
    Attention-UNet \cite{schlemper2019attention}  &  & 91.38 & 84.76 & 91.79 & 24.51 & & 76.89 & 62.92 & 77.45 & 21.63 & & 81.25 & 68.59 & 81.84 &  40.68 & & 84.25 & 75.91 & 85.75 & 54.07\\
    ResUNet++ \cite{jha2019resunet++} &  & 89.76 & 82.31 & 90.19 & 33.13 & & 77.96 & 64.05 & 78.85 & 19.45 & & 76.81 & 62.55 & 77.53 & 33.20 & & 83.06 & 73.52 & 84.52 & 62.47\\
    R2U-Net \cite{alom2019recurrent} & & 88.56 & 81.82 & 89.02 & 32.64 & & 78.87 & 65.32 &  79.35 & 21.34 & & 76.19 & 61.93 & 78.33 & 40.12 & & 83.79 & 74.67 & 85.49 &  58.11\\
    DoubleU-Net \cite{jha2020doubleu} &  & 91.09 & 84.43 & 91.62 & 27.47 & & 78.20 & 64.43 &   78.81 & 19.01 & & 81.59 & 68.98 &  81.83  & 40.66 & & 86.02 & 77.68 & 87.49 & 53.41\\
    UNet3+ \cite{huang2020unet} &  & 91.39 & 84.78  & 91.77 & 25.76 & & 78.03 & 64.25 & 78.46 & 19.25 & & 80.90 & 68.05 &  81.13  &  35.41 & & 84.49 & 75.76 & 86.11 & 55.72 \\
    TransUNet \cite{chen2021transunet} &  & 90.76 & 83.73 & 91.08 & 31.38 &  & 75.08 & 61.41 & 76.55 & 20.47 & & 78.30 & 64.37 & 77.65  &  40.94 & & 82.91 & 73.85 & 84.53 & 58.37\\
    nnU-Net \cite{isensee2021nnu} &  & 91.61 & 85.00 & 91.96 & 30.29 &  & 81.09 & 68.28 & 81.24 & 18.15 &  & 82.11 & 69.74 & 82.35 & 31.71 & & 87.35 & 79.32 & 88.48 & 52.09 \\
    \midrule
    Vanilla SAM \cite{kirillov2023_sam} & \multirow{5}{*}{\ding{56}} & 36.49 & 25.83 & 49.04 & 95.03 &  & 20.28 & 12.62 & 25.25 & 33.72 &  & 20.44 & 12.53 & 35.80 & 95.09 & & 39.88 & 19.17 & 40.40 & 201.56\\
    SAMMI \cite{huang2024segment} &  & 82.31 & 74.61 & 82.72 & 37.23 &  & 58.55 & 42.13 & 59.85 & 27.13 &  & 71.44 & 58.55 & 71.72 & 43.86 & & 75.74 & 64.37 & 77.52 & 87.71\\
    SAMUS \cite{lin2023samus} &  & 82.25 & 74.42 & 82.57 & 35.52 &  & 64.96 & 50.70 & 65.11 & 25.25 &  & 76.60 & 62.56 & 76.84 & 41.42 & & 77.45 & 66.65 & 79.22 & 90.47\\
    SAMed \cite{zhang2023customized_sam_liudong} & & 82.62 & 75.07 & 82.94 & 35.36 &  & 65.76 & 51.82 & 65.96 & 26.02  &  & 75.74 & 61.42 & 75.87 & 40.37 & & 77.96 & 67.22 & 79.74 & 86.72\\
    Med-SA \cite{wu2023medsa_jinyueming} &  & 82.84 & 75.24 & 83.14 & 34.72 &  & 64.91 & 51.02 & 65.06 & 25.25 &  & 77.59 & 62.95 & 78.63 & 40.65 & & 77.59 & 66.92 & 78.87 & 86.31\\
    \midrule
    Vanilla SAM \cite{kirillov2023_sam} & \multirow{5}{*}{point} & 70.72 & 59.10 & 74.56 & 44.03 & & 37.54 & 22.47 & 46.33 & 23.01 &  & 83.46 & 46.78 & 64.83 & 39.19 & & 62.67 & 44.78 & 72.74 & 154.98\\
    SAMMI \cite{huang2024segment} &  & 91.42 & 84.79 & 91.83 & 21.49 &  &  78.64 & 64.92 & 78.87 & 19.23 & & 80.45 & 67.43 & 80.58 & 30.91 & & 86.11 & 77.77 & 87.33 & 49.20 \\
    SAMUS \cite{lin2023samus} &  & 92.10 & 85.74 & 92.38 & 29.95 &  & 81.82 & 69.33 & 82.00 & 18.34 &  & 83.94 & 71.94 & 84.07 & 29.36 & & 87.52 & 79.85 & 88.70 & 53.69\\
    SAMed \cite{zhang2023customized_sam_liudong} &  & 92.39 & 86.27 & 92.66 & 27.20 &  & 82.26 & 69.94 & 82.43 & 18.64 &  & 83.19 & 71.23 & 83.33 & 33.82 & & 86.87 & 78.68 & 87.90 & 55.90\\
    Med-SA \cite{wu2023medsa_jinyueming} &  & 92.52 & 86.45 & 92.80 & 24.05 & & 82.47 & 70.24 & 82.57 & 18.21 &  & 84.16 & 72.68 & 84.32 & 29.11 & & 87.39 & 79.54 & 88.54 & 55.87\\
    \midrule
    UN-SAM (Ours)  & \ding{56} &  \textbf{93.12} & \textbf{87.41} & \textbf{93.30} & \textbf{20.90} &  &  \textbf{84.17} & \textbf{72.93} &  \textbf{84.27} & \textbf{16.55}  &  & \textbf{85.72} & \textbf{75.02} & \textbf{85.81} & \textbf{26.83} & & \textbf{89.01} & \textbf{82.14} & \textbf{89.88} & \textbf{46.15}\\
    \bottomrule
    \end{tabular}}}
    \label{tab_semantic_sota}
\end{table*}

\begin{table*}[t]
    \centering
    \small
    \setlength\tabcolsep{5pt}
    \caption{Comparison of Generalization Capability on Nuclei Semantic Segmentation.}
    {\scalebox{0.77}{
    \begin{tabular}{l|c|ccccccccccccccccccc}
    \toprule
    \multirow{2}{*}{Methods} & Manual & \multicolumn{4}{c}{DSB} & & \multicolumn{4}{c}{ MoNuSeg } & & \multicolumn{4}{c}{ TNBC } & & \multicolumn{4}{c}{ SegPC }\\
    \cline{3-6} \cline{8-11} \cline{13-16} \cline{18-21}
    & Prompt & Dice & mIoU & F1 & HD & & Dice & mIoU & F1 & HD & & Dice & mIoU & F1 & HD & & Dice & mIoU & F1 & HD\\
    \midrule
    Vanilla SAM \cite{kirillov2023_sam} & \multirow{5}{*}{\ding{56}} & 36.49 & 25.83 & 49.04 & 95.03 &  & 20.28 & 12.62 & 25.25 & 33.72 &  & 20.44 & 12.53 & 35.80 & 95.09 & & 39.88 & 19.17 & 40.40 & 201.56\\
    SAMMI \cite{huang2024segment} &  & 77.63 & 71.02 & 78.50 & 39.11 &  & 57.71 & 41.47 & 58.41 & 30.41 &  & 68.71 & 56.47 & 69.41 & 43.41 &  & 72.61 & 60.56 & 75.67 & 102.68\\
    SAMUS \cite{lin2023samus} & & 80.85 & 73.82 & 81.29 & 37.18 &  & 62.77 & 48.41 & 75.10 & 26.07 &  & 74.18 & 60.56 & 74.79 & 40.24 &  & 73.47 & 61.91 & 75.38 & 86.21\\
    SAMed \cite{zhang2023customized_sam_liudong} &  & 80.87 & 73.85 & 81.23 & 37.73 &  & 64.42 & 50.61 & 65.54 & 25.14 & & 73.51 & 59.67 & 73.81 & 40.73 & & 75.47 & 64.46 & 77.46 & 80.21 \\
    Med-SA \cite{wu2023medsa_jinyueming} &  & 81.54 & 74.78 & 81.84 & 36.07 &  & 63.57 & 49.15 & 64.83 & 25.96 & & 74.27 & 60.49 & 75.42 & 39.42 &  & 74.70 & 62.24 & 75.92 & 81.07\\
    \midrule
    Vanilla SAM \cite{kirillov2023_sam} & \multirow{5}{*}{point} & 70.72 & 59.10 & 74.56 & 44.03 & & 37.54 & 22.47 & 46.33 & 23.01 &  & 83.46 & 46.78 & 64.83 & 39.19 & & 62.67 & 44.78 & 72.74 & 154.98\\
    SAMMI\cite{huang2024segment} &  & 90.37 & 83.15 & 90.77 & 33.67 &  & 77.70 & 63.80 & 78.07 & 19.61 &  & 77.77 & 63.83 & 78.19 & 38.93 &  & 85.19 & 76.36 & 85.19 & 67.43\\
    SAMUS \cite{lin2023samus} &  & 91.39 & 84.69 & 91.76 & 29.45 &  & 78.11 & 64.22 & 78.41 & 18.75 &  & 81.28 & 68.48 & 81.42 & 33.18 &  & 86.49 & 78.32 & 87.75 & 55.25\\
    SAMed \cite{zhang2023customized_sam_liudong} &  & 91.49 & 84.89 & 91.83 & 30.92 &  & 80.14 & 66.96 & 80.37 & 18.78 &  & 81.69 & 66.74 & 80.22 & 32.43 &  & 86.55 & 78.37 & 87.67 & 53.98\\  
    Med-SA \cite{wu2023medsa_jinyueming} &  & 92.16 & 85.99 & 92.46 & 25.35 & & 80.79 & 67.87 & 81.01 & 17.68 &  & 81.99 & 69.58 & 82.29 & 34.65 &  & 86.13 & 77.79 & 87.56 & 49.26\\
    \midrule
    UN-SAM (Ours) & \ding{56} & \textbf{92.84} & \textbf{86.93} & \textbf{93.03} & \textbf{20.93} &  &  \textbf{82.86} & \textbf{70.82} &  \textbf{82.96} & \textbf{16.69}  &  & \textbf{83.89} & \textbf{72.27} & \textbf{84.05} & \textbf{27.71} & & \textbf{88.17} & \textbf{81.06} & \textbf{89.05} & \textbf{48.93}\\
    \bottomrule
    \end{tabular}}}
    \label{tab_semantic_generalized}
\end{table*}

\subsection{Comparison on Nuclei Semantic Segmentation} \label{sec:css}
To perform the universal evaluation of our UN-SAM in nuclei segmentation, we further conduct comparisons on nuclei semantic segmentation with advanced medical segmentation algorithms. First, we compare the performance of models fine-tuned on each single nuclei domain, as shown in Table \ref{tab_semantic_sota}. Among medical segmentation methods, nnU-Net \cite{isensee2021nnu} achieves leading results in nuclei semantic segmentation tasks, and also outperforms fine-tuned SAM methods when manual prompts are unavailable. In contrast, these fine-tuned SAM methods \cite{huang2024segment,lin2023samus,zhang2023customized_sam_liudong,wu2023medsa_jinyueming} benefit from manual prompts and reveal remarkable performance over classical medical segmentation methods, \textit{e.g.}, Med-SA \cite{wu2023medsa_jinyueming} has a 2.05\% Dice increase over nnU-Net \cite{isensee2021nnu} on the TNBC dataset. Our UN-SAM further promotes SAM capability on nuclei segmentation tasks, and achieves overwhelming performance on these four datasets, with the best Dice of 93.12\%, 84.17\%, 85.72\% and 89.01\%, respectively.

\begin{figure*}[!t]
  \centering
  \includegraphics[width=0.85\linewidth]{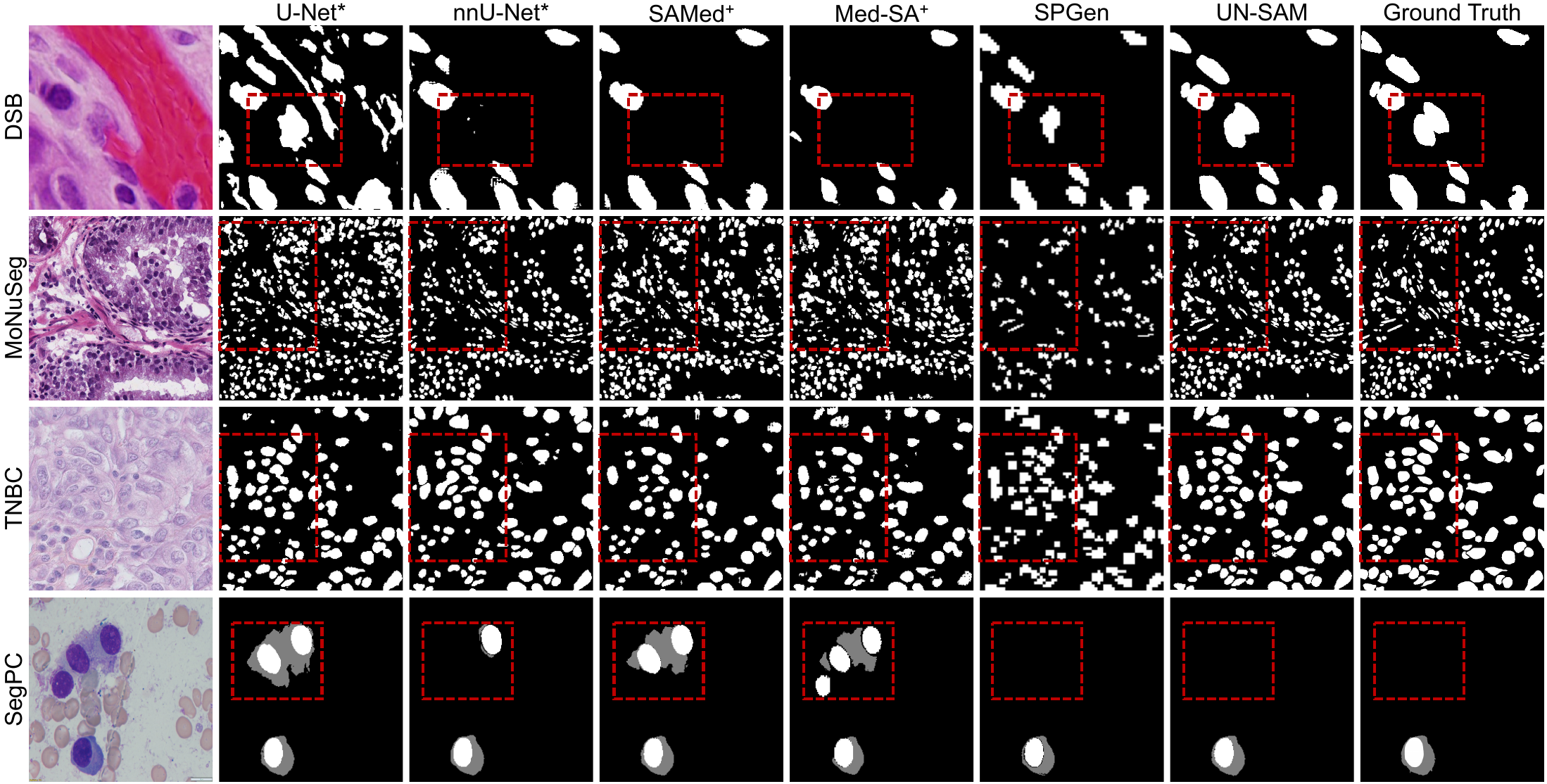}
  \caption{Visualization of generalized nuclei semantic segmentation. ${+}$ indicates medical SAMs using point prompts, and $*$ indicates that classical methods are trained and tested separately for each dataset. Our UN-SAM exhibits the best results, segmenting more nuclei with accurate boundaries while having fewer false positives. The SPGen produces the self-prompt to provide high-quality segmentation hints for our UN-SAM.}
  \label{fig:visual_semantic}
\end{figure*}

Moreover, we compare the generalization capability of SAM-based methods across different nuclei domains after fine-tuning them on these four datasets sequentially, as shown in Table \ref{tab_semantic_generalized}. Our UN-SAM achieves the best performance on these four domains of nuclei images simultaneously, which confirms the advantage of generalization capability in semantic segmentation tasks. In particular, our UN-SAM surpasses the second-best Med-SA \cite{wu2023medsa_jinyueming} without manual prompts, by a remarkable Dice increase of 11.30\%, 19.29\%, 9.62\% and 13.47\% on these four datasets, respectively. Compared with point-prompted medical SAMs, our UN-SAM reveals more than 2\% advantage of Dice in most domains. We further elaborate on the semantic segmentation results in Fig~\ref{fig:visual_semantic}, and these four datasets reveal significant data heterogeneity due to different tissue types, staining protocols, and imaging conditions. Our UN-SAM generates the best segmentation results, especially reducing false positive predictions of nuclei significantly. These comparisons validate the superiority of our UN-SAM on nuclei semantic segmentation tasks and better generalization performance across different nuclei domains.

\begin{figure}[t]
  \centering
  \includegraphics[width=0.99\linewidth]{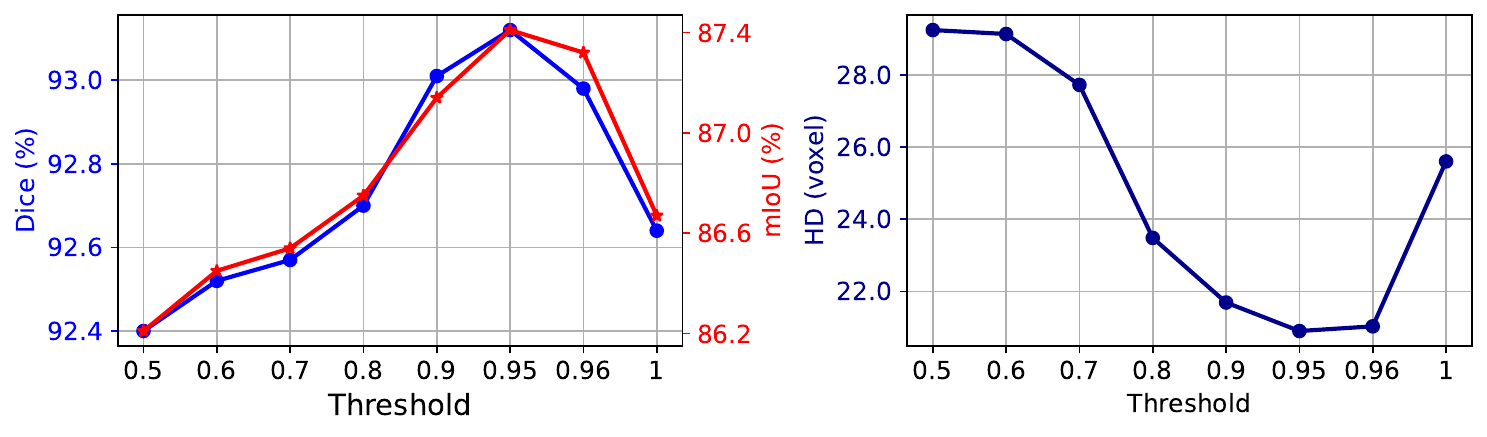}
  \caption{Hyper-parameter analysis of confidence threshold on nuclei semantic segmentation of the DSB dataset.}
  \label{fig:hyper}
\end{figure}

\begin{table*}[t]
    \centering
    \caption{Ablation Study of UN-SAM on Nuclei Semantic Segmentation.}
    {\scalebox{0.82}{
    \begin{tabular}{cccc|cccccccccccccccccccc}
    \toprule
     & \multirow{2}{*}{$\bm{E}$} & \multirow{2}{*}{$\bm{S}$} & \multirow{2}{*}{$\bm{D}$}  & \multicolumn{4}{c}{DSB} & & \multicolumn{4}{c}{ MoNuSeg } & & \multicolumn{4}{c}{ TNBC } & & \multicolumn{4}{c}{ SegPC-2021 }\\
    \cline{5-8} \cline{10-13} \cline{15-18} \cline{20-23}
    & &  &  & Dice & mIoU & F1 & HD & & Dice & mIoU & F1 & HD & & Dice & mIoU & F1 & HD & & Dice & mIoU & F1 & HD\\
    \midrule
    1 & &  &  &  90.08 & 82.62 & 90.59 & 36.82 &  & 72.43 & 57.15 & 73.13 & 18.56 &  & 76.83 & 62.65 & 77.54 & 36.15 &  & 73.47 & 61.91 & 75.38 & 86.21 \\
    2 & \checkmark &  & & 92.14 & 85.96 & 92.46 & 30.98 & & 82.27 & 70.02 & 82.41 & 17.41 & & 82.75 & 70.61 & 82.84 & 31.25 & & 87.20 & 79.27 & 88.34 & 54.79\\
    3 & & \checkmark &  & 90.18 & 82.77 & 90.72 & 31.74 & & 76.47 & 62.04 & 76.69 & 17.62 & & 78.92 & 65.32 & 79.32 & 33.13 & & 74.95 & 63.35 & 76.86 & 76.66\\
    4 & &  & \checkmark & 91.90 & 85.38 & 92.21 & 28.23 & & 82.19 & 69.85 & 82.38 & 17.37  & & 82.57 & 70.34 & 82.69 & 31.59 & & 86.37 & 78.12 & 87.73 & 55.62\\
    5 & \checkmark & &  \checkmark & 92.63 & 86.65 & 92.87 & 25.51 & & 82.73 & 70.53 & 82.81 & 16.94 & & 83.33 & 71.48 & 83.41 & 30.12 & & 88.00 & 81.02 & 89.03 & 51.37\\
    6 & \checkmark & \checkmark & & 92.44 & 86.32 & 92.69 & 21.73 & & 82.42 & 70.12 & 82.50 & 16.78 & & 83.02 & 70.98 & 83.24 & 28.16 & & 87.89 & 79.91 & 88.82 & 49.25\\
    7 & & \checkmark & \checkmark & 92.19 & 85.92 & 92.53 & 24.44 & & 82.22 & 69.88 & 82.38 & 17.06 & & 82.71 & 70.51 & 82.80 & 29.11 & & 87.20 & 79.10 & 88.28 & 49.61\\
    8 & \checkmark & \checkmark & \checkmark & \textbf{92.84} & \textbf{86.93} & \textbf{93.03} & \textbf{20.93} &  &  \textbf{82.86} & \textbf{70.82} &  \textbf{82.96} & \textbf{16.69}  &  & \textbf{83.89} & \textbf{72.27} & \textbf{84.05} & \textbf{27.71} & & \textbf{88.17} & \textbf{81.06} & \textbf{89.05} & \textbf{48.93}\\
  \bottomrule
  \end{tabular}}}
  \label{tab_semantic_ablation}
\end{table*}

\subsection{Ablation Study}

To investigate the effectiveness of the DT-Encoder $\bm{E}$, SPGen module $\bm{S}$ and DQ-Decoder $\bm{D}$, we further conduct the comprehensive ablation study on the semantic segmentation of four nuclei datasets, as illustrated in Table \ref{tab_semantic_ablation}. By removing the tailored modules from UN-SAM, the fine-tuned SAM ($1^{st}$ row) serves as the ablation baseline. By separately introducing the DT-Encoder ($2^{nd}$ row), SPGen ($3^{rd}$ row) and DQ-Decoder ($4^{th}$ row), the performance is improved with the average Dice of all datasets gain of 7.89\%, 1.93\%, 7.56\%, respectively. We additionally investigate the effect of combined DT-Encoder and DQ-Decoder ($5^{th}$ row), resulting in superior performance, with the Dice of 92.63\%, 82.73\%, 83.33\%, 88.00\% and mIoU of 86.65\%, 70.53\%, 71.48\%, 81.02\% on four datasets. The result proves that these two improvements can promote the generalization across different nuclei domains. By comparing $6^{th}$ and $7^{th}$ rows with $2^{nd}$ and $4^{th}$ rows, configurations with SPGen show competitive performance compared to the manual prompt encoder of SAM, which overcomes the laborious process of manual prompt generation and facilitates clinical workflows for high-throughput analysis. On this basis, our UN-SAM ($8^{th}$ row) simultaneously adopts DT-Encoder and DQ-Decoder to further address both issues. In this way, these ablation experiments prove that the tailor-made DT-Encoder, SPGen module, and DQ-Decoder play a significant role to facilitate our UN-SAM on nuclei segmentation by eliminating the annotation need and enhancing the generalization across varying nuclei domains.

It is worth noting that the confidence threshold $\tau$ in the SPGen module is crucial for UN-SAM, by determining the high-quality self-prompt to guide the segmentation decoding. For the possible value range as $0.5 \leq \tau \leq 1$, we perform a grid search with the Dice, mIoU and HD metrics on the DSB dataset. As shown in Fig. \ref{fig:hyper}, our UN-SAM achieves the best performance when $\tau$ is 0.95, where a lower threshold would cause interference to the decoder and a larger threshold cannot provide enough prompt information for nuclei segmentation. In general, the proposed SPGen module with the appropriate confidence threshold can significantly benefit the UN-SAM and eliminate the dependence on manual annotation for segmentation prompts.

\begin{table}[t]
    \centering
    \small
    \setlength\tabcolsep{5pt}
    \caption{Zero-shot Generalization Comparison for CryoNuSeg.}
    {\scalebox{0.85}{
    \begin{tabular}{l|c|c|cccc}
    \toprule
    \multirow{2}{*}{Methods} & \multirow{2}{*}{Mode} & Manual & \multirow{2}{*}{Dice} & \multirow{2}{*}{mIoU} & \multirow{2}{*}{F1} & \multirow{2}{*}{HD} \\
     & & Prompt &  &  &  &  \\ 
    \midrule

    U-Net \cite{ronneberger2015u}& \multirow{2}{*}{fine-tune} & \multirow{2}{*}{\ding{56}}  & 80.15 & 67.22 & 80.19 & 25.24 \\
    nnU-Net \cite{isensee2021nnu} & & & 82.15 & 69.56 & 82.22 & 23.15 \\

    \midrule
    
    Vanilla SAM \cite{kirillov2023_sam} & \multirow{5}{*}{zero-shot} & \multirow{5}{*}{\ding{56}}& 40.86 & 26.19 & 43.87 & 86.13 \\
    SAMMI \cite{huang2024segment} &  & & 64.14 & 50.73 & 65.19 & 37.88\\
    SAMUS \cite{lin2023samus}&  &   & 65.52 & 51.01 & 66.63 & 35.43 \\
    SAMed \cite{zhang2023customized_sam_liudong} &  & & 67.01 & 53.53 & 68.61 & 36.00 \\
    Med-SA \cite{wu2023medsa_jinyueming} &  & & 67.16 & 53.60 & 68.82 & 34.36\\
    \midrule
    Vanilla SAM \cite{kirillov2023_sam} & \multirow{5}{*}{zero-shot} & \multirow{5}{*}{point} & 61.05 & 44.58 & 66.29 & 30.51 \\
    SAMMI \cite{huang2024segment} &  & & 75.49 & 61.39 & 76.25 & 29.66 \\
    SAMUS \cite{lin2023samus} &  &  & 77.78 & 64.36 & 78.50 & 25.32 \\
    SAMed \cite{zhang2023customized_sam_liudong} &  & & 78.20 & 64.85 & 78.93 & 25.29\\
    Med-SA \cite{wu2023medsa_jinyueming} & &  & 78.46 & 65.18 & 79.10 & 25.80 \\
    \midrule
    UN-SAM (Ours) & zero-shot & \ding{56} & \textbf{80.42} & \textbf{67.67} & \textbf{80.67} & \textbf{25.18} \\
    
    \bottomrule
    \end{tabular}}}
    \label{tab_semantic_zeroshot}
\end{table}

\subsection{Comparison on Zero-shot Generalization}
To further validate the generalization capability of our UN-SAM, we compare zero-shot generalization of medical foundation models (\textit{i.e.}, the model weights in Table \ref{tab_semantic_generalized}) on the CryoNuSeg \cite{mahbod2021cryonuseg} test set. As demonstrated in Table \ref{tab_semantic_zeroshot}, our UN-SAM achieves the best performance among medical SAMs, with the overwhelming Dice of 80.42\% and mIoU of 67.67\%. In particular, our UN-SAM reveals significant performance advantages with the P-value $<$ 0.005 compared to all medical SAMs \cite{huang2024segment,lin2023samus,zhang2023customized_sam_liudong,wu2023medsa_jinyueming} with or without manual prompts. Moreover, we present medical segmentation methods, \textit{i.e.}, U-Net \cite{ronneberger2015u} and nnU-Net \cite{isensee2021nnu} fine-tuned on the CryoNuSeg training set. Remarkably, our UN-SAM outperforms the fine-tuned U-Net \cite{ronneberger2015u}, and is approaching the fined-tuned nnU-Net \cite{isensee2021nnu} as the upper bound. These comparisons confirm that our UN-SAM has superior zero-shot generalization capability, with more potential to be applied in clinical scenarios.

\section{Conclusion}
In this work, we propose the UN-SAM framework to achieve a universal solution for nuclei segmentation within the realm of digital pathology. In particular, the DT-Encoder is proposed to harmonize domain-common and domain-specific features within the image encoder, which propels the generalization potential of SAM across diverse nuclei datasets. Then, the SPGen module is devised to autonomously produce high-quality mask hints instead of manual annotated prompt, to facilitate the clinical workflow towards enhanced efficiency. Moreover, we further propose the DQ-Decoder to leverage learnable domain queries for segmentation decoding in different nuclei domains. Extensive experiments confirm the advantage of UN-SAM over existing medical SAMs by negating the need for labor-intensive manual annotations and offering an unparalleled zero-shot learning capability that adapts to varying domains with ease.

\balance
\bibliographystyle{IEEEtran}
\bibliography{czmybibliography}

\begin{thebibliography}{10}
\providecommand{\url}[1]{#1}
\csname url@samestyle\endcsname
\providecommand{\newblock}{\relax}
\providecommand{\bibinfo}[2]{#2}
\providecommand{\BIBentrySTDinterwordspacing}{\spaceskip=0pt\relax}
\providecommand{\BIBentryALTinterwordstretchfactor}{4}
\providecommand{\BIBentryALTinterwordspacing}{\spaceskip=\fontdimen2\font plus
\BIBentryALTinterwordstretchfactor\fontdimen3\font minus \fontdimen4\font\relax}
\providecommand{\BIBforeignlanguage}[2]{{%
\expandafter\ifx\csname l@#1\endcsname\relax
\typeout{** WARNING: IEEEtran.bst: No hyphenation pattern has been}%
\typeout{** loaded for the language `#1'. Using the pattern for}%
\typeout{** the default language instead.}%
\else
\language=\csname l@#1\endcsname
\fi
#2}}
\providecommand{\BIBdecl}{\relax}
\BIBdecl

\bibitem{greenwald2022whole}
N.~F. Greenwald, G.~Miller, E.~Moen, A.~Kong, A.~Kagel, T.~Dougherty, C.~C. Fullaway, B.~J. McIntosh, K.~X. Leow, M.~S. Schwartz \emph{et~al.}, ``Whole-cell segmentation of tissue images with human-level performance using large-scale data annotation and deep learning,'' \emph{Nature Biotechnology}, vol.~40, no.~4, pp. 555--565, 2022.

\bibitem{caicedo2019nucleus}
J.~C. Caicedo, A.~Goodman, K.~W. Karhohs, B.~A. Cimini, J.~Ackerman, M.~Haghighi, C.~Heng, T.~Becker, M.~Doan, C.~McQuin \emph{et~al.}, ``Nucleus segmentation across imaging experiments: the 2018 data science bowl,'' \emph{Nature Methods}, vol.~16, no.~12, pp. 1247--1253, 2019.

\bibitem{gupta2023segpc}
A.~Gupta, S.~Gehlot, S.~Goswami, S.~Motwani, R.~Gupta, {\'A}.~G. Faura, D.~{\v{S}}tepec, T.~Martin{\v{c}}i{\v{c}}, R.~Azad, D.~Merhof \emph{et~al.}, ``Segpc-2021: A challenge \& dataset on segmentation of multiple myeloma plasma cells from microscopic images,'' \emph{Med. Image Anal.}, vol.~83, p. 102677, 2023.

\bibitem{he2017mask}
K.~He, G.~Gkioxari, P.~Doll{\'a}r, and R.~Girshick, ``Mask r-cnn,'' in \emph{ICCV}, 2017, pp. 2961--2969.

\bibitem{cheng2022sparse}
T.~Cheng, X.~Wang, S.~Chen, W.~Zhang, Q.~Zhang, C.~Huang, Z.~Zhang, and W.~Liu, ``Sparse instance activation for real-time instance segmentation,'' in \emph{CVPR}, 2022, pp. 4433--4442.

\bibitem{graham2019hover}
S.~Graham, Q.~D. Vu, S.~E.~A. Raza, A.~Azam, Y.~W. Tsang, J.~T. Kwak, and N.~Rajpoot, ``Hover-net: Simultaneous segmentation and classification of nuclei in multi-tissue histology images,'' \emph{Med. Image Anal.}, vol.~58, p. 101563, 2019.

\bibitem{chen2023cpp}
S.~Chen, C.~Ding, M.~Liu, J.~Cheng, and D.~Tao, ``Cpp-net: Context-aware polygon proposal network for nucleus segmentation,'' \emph{IEEE Trans. Image Process.}, vol.~32, pp. 980--994, 2023.

\bibitem{he2021cdnet}
H.~He, Z.~Huang, Y.~Ding, G.~Song, L.~Wang, Q.~Ren, P.~Wei, Z.~Gao, and J.~Chen, ``Cdnet: Centripetal direction network for nuclear instance segmentation,'' in \emph{ICCV}, 2021, pp. 4026--4035.

\bibitem{nam2023pronet}
S.~Nam, J.~Jeong, M.~Luna, P.~Chikontwe, and S.~H. Park, ``Pronet: Point refinement using shape-guided offset map for nuclei instance segmentation,'' in \emph{MICCAI}.\hskip 1em plus 0.5em minus 0.4em\relax Springer, 2023, pp. 528--538.

\bibitem{he2023toposeg}
H.~He, J.~Wang, P.~Wei, F.~Xu, X.~Ji, C.~Liu, and J.~Chen, ``Toposeg: Topology-aware nuclear instance segmentation,'' in \emph{ICCV}, 2023, pp. 21\,307--21\,316.

\bibitem{kirillov2023_sam}
A.~Kirillov, E.~Mintun, N.~Ravi, H.~Mao, C.~Rolland, L.~Gustafson, T.~Xiao, S.~Whitehead, A.~C. Berg, W.-Y. Lo, P.~Dollar, and R.~Girshick, ``Segment anything,'' in \emph{ICCV}, October 2023, pp. 4015--4026.

\bibitem{cheng2023sam_med2d}
J.~Cheng, J.~Ye, Z.~Deng, J.~Chen, T.~Li, H.~Wang, Y.~Su, Z.~Huang, J.~Chen, L.~Jiang \emph{et~al.}, ``Sam-med2d,'' \emph{arXiv preprint arXiv:2308.16184}, 2023.

\bibitem{zhang2024segment_sam_survery}
Y.~Zhang, Z.~Shen, and R.~Jiao, ``Segment anything model for medical image segmentation: Current applications and future directions,'' \emph{arXiv preprint arXiv:2401.03495}, 2024.

\bibitem{ma2024sam_majun_nc}
J.~Ma, Y.~He, F.~Li, L.~Han, C.~You, and B.~Wang, ``Segment anything in medical images,'' \emph{Nature Communications}, vol.~15, no.~1, p. 654, 2024.

\bibitem{huang2024segment}
Y.~Huang, X.~Yang, L.~Liu, H.~Zhou, A.~Chang, X.~Zhou, R.~Chen, J.~Yu, J.~Chen, C.~Chen \emph{et~al.}, ``Segment anything model for medical images?'' \emph{Med. Image Anal.}, vol.~92, p. 103061, 2024.

\bibitem{lin2023samus}
X.~Lin, Y.~Xiang, L.~Zhang, X.~Yang, Z.~Yan, and L.~Yu, ``Samus: Adapting segment anything model for clinically-friendly and generalizable ultrasound image segmentation,'' \emph{arXiv preprint arXiv:2309.06824}, 2023.

\bibitem{zhang2023customized_sam_liudong}
K.~Zhang and D.~Liu, ``Customized segment anything model for medical image segmentation,'' \emph{arXiv preprint arXiv:2304.13785}, 2023.

\bibitem{wu2023medsa_jinyueming}
J.~Wu, W.~Ji, Y.~Liu, H.~Fu, M.~Xu, Y.~Xu, and Y.~Jin, ``Medical sam adapter: Adapting segment anything model for medical image segmentation,'' \emph{arXiv preprint arXiv:2304.12620}, 2023.

\bibitem{xu2023sppnet}
Q.~Xu, W.~Kuang, Z.~Zhang, X.~Bao, H.~Chen, and W.~Duan, ``Sppnet: A single-point prompt network for nuclei image segmentation,'' in \emph{MICCAI Workshop on MLMI}.\hskip 1em plus 0.5em minus 0.4em\relax Springer, 2023, pp. 227--236.

\bibitem{isensee2021nnu}
F.~Isensee, P.~F. Jaeger, S.~A. Kohl, J.~Petersen, and K.~H. Maier-Hein, ``nnu-net: a self-configuring method for deep learning-based biomedical image segmentation,'' \emph{Nature Methods}, vol.~18, no.~2, pp. 203--211, 2021.

\bibitem{ronneberger2015u}
O.~Ronneberger, P.~Fischer, and T.~Brox, ``U-net: Convolutional networks for biomedical image segmentation,'' in \emph{MICCAI}.\hskip 1em plus 0.5em minus 0.4em\relax Springer, 2015, pp. 234--241.

\bibitem{alom2019recurrent}
M.~Z. Alom, C.~Yakopcic, M.~Hasan, T.~M. Taha, and V.~K. Asari, ``Recurrent residual u-net for medical image segmentation,'' \emph{J. Med. Imaging}, vol.~6, no.~1, pp. 014\,006--014\,006, 2019.

\bibitem{chen2021transunet}
J.~Chen, Y.~Lu, Q.~Yu, X.~Luo, E.~Adeli, Y.~Wang, L.~Lu, A.~L. Yuille, and Y.~Zhou, ``Transunet: Transformers make strong encoders for medical image segmentation,'' \emph{arXiv preprint arXiv:2102.04306}, 2021.

\bibitem{schlemper2019attention}
J.~Schlemper, O.~Oktay, M.~Schaap, M.~Heinrich, B.~Kainz, B.~Glocker, and D.~Rueckert, ``Attention gated networks: Learning to leverage salient regions in medical images,'' \emph{Med. Image Anal.}, vol.~53, pp. 197--207, 2019.

\bibitem{zhou2019unet++}
Z.~Zhou, M.~M.~R. Siddiquee, N.~Tajbakhsh, and J.~Liang, ``Unet++: Redesigning skip connections to exploit multiscale features in image segmentation,'' \emph{IEEE Trans. Med. Imaging}, vol.~39, no.~6, pp. 1856--1867, 2019.

\bibitem{lin2022contrans}
A.~Lin, J.~Xu, J.~Li, and G.~Lu, ``Contrans: Improving transformer with convolutional attention for medical image segmentation,'' in \emph{MICCAI}.\hskip 1em plus 0.5em minus 0.4em\relax Springer, 2022, pp. 297--307.

\bibitem{kumar2017dataset}
N.~Kumar, R.~Verma, S.~Sharma, S.~Bhargava, A.~Vahadane, and A.~Sethi, ``A dataset and a technique for generalized nuclear segmentation for computational pathology,'' \emph{IEEE Trans. Med. Imaging}, vol.~36, no.~7, pp. 1550--1560, 2017.

\bibitem{naylor2018segmentation}
P.~Naylor, M.~La{\'e}, F.~Reyal, and T.~Walter, ``Segmentation of nuclei in histopathology images by deep regression of the distance map,'' \emph{IEEE Trans. Med. Imaging}, vol.~38, no.~2, pp. 448--459, 2018.

\bibitem{xiong2023efficientsam}
Y.~Xiong, B.~Varadarajan, L.~Wu, X.~Xiang, F.~Xiao, C.~Zhu, X.~Dai, D.~Wang, F.~Sun, F.~Iandola \emph{et~al.}, ``Efficientsam: Leveraged masked image pretraining for efficient segment anything,'' \emph{arXiv preprint arXiv:2312.00863}, 2023.

\bibitem{shui2023unleashing_sam_yanglin}
Z.~Shui, Y.~Zhang, K.~Yao, C.~Zhu, Y.~Sun, and L.~Yang, ``Unleashing the power of prompt-driven nucleus instance segmentation,'' \emph{arXiv preprint arXiv:2311.15939}, 2023.

\bibitem{hu2022lora}
E.~J. Hu, Y.~Shen, P.~Wallis, Z.~Allen-Zhu, Y.~Li, S.~Wang, L.~Wang, and W.~Chen, ``Lo{RA}: Low-rank adaptation of large language models,'' in \emph{ICLR}, 2022.

\bibitem{houlsby2019parameter}
N.~Houlsby, A.~Giurgiu, S.~Jastrzebski, B.~Morrone, Q.~De~Laroussilhe, A.~Gesmundo, M.~Attariyan, and S.~Gelly, ``Parameter-efficient transfer learning for nlp,'' in \emph{ICML}.\hskip 1em plus 0.5em minus 0.4em\relax PMLR, 2019, pp. 2790--2799.

\bibitem{anonymous2024convolution}
Z.~Zhong, Z.~Tang, T.~He, H.~Fang, and C.~Yuan, ``Convolution meets lora: Parameter efficient finetuning for segment anything model,'' in \emph{ICLR}, 2024.

\bibitem{jieboluo2024surgicalsam}
W.~Yue, J.~Zhang, K.~Hu, Y.~Xia, J.~Luo, and Z.~Wang, ``Surgicalsam: Efficient class promptable surgical instrument segmentation,'' in \emph{AAAI}, 2024.

\bibitem{cui2023all_in_sam}
C.~Cui, R.~Deng, Q.~Liu, T.~Yao, S.~Bao, L.~W. Remedios, Y.~Tang, and Y.~Huo, ``All-in-sam: from weak annotation to pixel-wise nuclei segmentation with prompt-based finetuning,'' \emph{arXiv preprint arXiv:2307.00290}, 2023.

\bibitem{dosovitskiy2020image}
A.~Dosovitskiy, L.~Beyer, A.~Kolesnikov, D.~Weissenborn, X.~Zhai, T.~Unterthiner, M.~Dehghani, M.~Minderer, G.~Heigold, S.~Gelly \emph{et~al.}, ``An image is worth 16x16 words: Transformers for image recognition at scale,'' in \emph{ICLR}, 2020.

\bibitem{ali2021xcit}
A.~Ali, H.~Touvron, M.~Caron, P.~Bojanowski, M.~Douze, A.~Joulin, I.~Laptev, N.~Neverova, G.~Synnaeve, J.~Verbeek \emph{et~al.}, ``Xcit: Cross-covariance image transformers,'' \emph{NeurIPS}, vol.~34, pp. 20\,014--20\,027, 2021.

\bibitem{mahbod2021cryonuseg}
A.~Mahbod, G.~Schaefer, B.~Bancher, C.~L{\"o}w, G.~Dorffner, R.~Ecker, and I.~Ellinger, ``Cryonuseg: A dataset for nuclei instance segmentation of cryosectioned h\&e-stained histological images,'' \emph{Comput. Biol. Med.}, vol. 132, p. 104349, 2021.

\bibitem{jha2019resunet++}
D.~Jha, P.~H. Smedsrud, M.~A. Riegler, D.~Johansen, T.~De~Lange, P.~Halvorsen, and H.~D. Johansen, ``Resunet++: An advanced architecture for medical image segmentation,'' in \emph{ISM}.\hskip 1em plus 0.5em minus 0.4em\relax IEEE, 2019, pp. 225--2255.

\bibitem{jha2020doubleu}
D.~Jha, M.~A. Riegler, D.~Johansen, P.~Halvorsen, and H.~D. Johansen, ``Doubleu-net: A deep convolutional neural network for medical image segmentation,'' in \emph{CBMS}.\hskip 1em plus 0.5em minus 0.4em\relax IEEE, 2020, pp. 558--564.

\bibitem{huang2020unet}
H.~Huang, L.~Lin, R.~Tong, H.~Hu, Q.~Zhang, Y.~Iwamoto, X.~Han, Y.-W. Chen, and J.~Wu, ``Unet 3+: A full-scale connected unet for medical image segmentation,'' in \emph{ICASSP}.\hskip 1em plus 0.5em minus 0.4em\relax IEEE, 2020, pp. 1055--1059.

\end{thebibliography}

\end{document}